\documentclass[pdflatex,sn-nature]{sn-jnl}

\usepackage{amsmath,amssymb,bm}
\usepackage{booktabs}
\usepackage{array}
\newcolumntype{L}[1]{>{\raggedright\arraybackslash}p{#1}}
\usepackage{graphicx}
\usepackage{placeins}
\usepackage{xcolor}
\usepackage{soul}

\graphicspath{{figs/}}
\newcommand{\dev}{\operatorname{dev}}
\newcommand{\tr}{\operatorname{tr}}
\newcommand{\I}{\mathbf I}
\newcommand{\bL}{\mathbf L}
\newcommand{\bG}{\mathbf G}
\newcommand{\bH}{\mathbf H}
\newcommand{\bu}{\mathbf u}
\newcommand{\br}{\mathbf r}
\newcommand{\bq}{\mathbf q}

\newcommand{\bh}{\mathbf h}
\newcommand{\bx}{\mathbf x}
\newcommand{\bA}{\mathbf A}

\newcommand{\bT}{\mathbf T}
\newcommand{\gam}{\bm\gamma}
\newcommand{\hatr}{\hat{\mathbf r}}
\newcommand{\bE}{\mathbf E}
\newcommand{\shp}{\mathrm S}

\newcommand{\Nbench}{18{,}056}

\newcommand{\Nneutron}{28}

\newcommand{\Ndev}{7{,}765}
\newcommand{\Nomat}{124}
\newcommand{\Ntestmin}{1{,}806}

\newcommand{\Ntrainmin}{14{,}625}

\newcommand{\Nvalmin}{1{,}625}

\newcommand{\Nrows}{17{,}937}

\newcommand{\NrowsB}{17{,}641}

\newcommand{\eVPBE}{4.12}
\newcommand{\eVFin}{0.85}

\newcommand{\eSPBE}{0.93}
\newcommand{\eSFin}{0.64}

\newcommand{\eBPBE}{1.65}
\newcommand{\eBIso}{0.99}
\newcommand{\eBMeas}{0.96}

\newcommand{\eBFin}{0.86}
\newcommand{\eDPBE}{0.043}

\newcommand{\eDFin}{0.035}
\newcommand{\eTPBE}{0.91}

\newcommand{\eTFin}{0.78}
\newcommand{\eVPBEmed}{4.05}
\newcommand{\eVFinmed}{0.48}
\newcommand{\eSPBEmed}{0.54}
\newcommand{\eSFinmed}{0.36}
\newcommand{\eBPBEmed}{1.48}
\newcommand{\eBFinmed}{0.51}
\newcommand{\eSnoD}{0.67}
\newcommand{\eVnoD}{0.89}

\newcommand{\impS}{73}

\newcommand{\bandshare}{70}

\newcommand{\bandworse}{36}

\newcommand{\tailPBE}{4.60}
\newcommand{\tailFin}{3.02}

\newcommand{\eSone}{0.65}
\newcommand{\eVone}{0.86}

\newcommand{\XHn}{19}
\newcommand{\XHPBE}{3.20}

\newcommand{\XHFin}{2.38}

\newcommand{\redV}{79}
\newcommand{\redS}{31}
\newcommand{\redB}{48}

\newcommand{\constVdft}{2.08}

\newcommand{\constXHdft}{2.72}

\newcommand{\redteceV}{81}
\newcommand{\redteceS}{40}
\newcommand{\redteceB}{49}

\newcommand{\redeqfV}{79}
\newcommand{\redeqfS}{39}
\newcommand{\redeqfB}{48}

\newcommand{\randVrello}{-80}
\newcommand{\randVrelhi}{-78}

\newcommand{\randSrello}{-32}
\newcommand{\randSrelhi}{-29}

\newcommand{\randBrello}{-49}
\newcommand{\randBrelhi}{-47}

\newcommand{\expteceOutn}{5{,}307}

\newcommand{\gPVcorr}{0.89}

\newcommand{\gPScorr}{0.68}

\newcommand{\gPBcorr}{0.89}

\newcommand{\gCVcorr}{0.85}

\newcommand{\gCScorr}{0.65}

\newcommand{\gCBcorr}{0.86}

\newcommand{\tailpopshare}{5.8}
\newcommand{\tailerrshare}{30}
\newcommand{\tailid}{7}

\newcommand{\tailcodn}{141}
\newcommand{\tailcod}{0.19}
\newcommand{\tailcodpbe}{4.23}
\newcommand{\tailrscann}{415}
\newcommand{\tailrscan}{0.998}

\newcommand{\shapesubn}{1{,}450}
\newcommand{\shapesubIR}{0.10}
\newcommand{\shapesubPBE}{0.37}
\newcommand{\shapesubFin}{0.25}

\newcommand{\codsubn}{1{,}428}
\newcommand{\codsubIR}{0.55}

\newcommand{\codsubPBE}{1.30}
\newcommand{\codsubFin}{0.43}

\newcommand{\tailnonthreeD}{26}
\newcommand{\allnonthreeD}{10}

\newcommand{\sgdiffshare}{2.4}
\newcommand{\sgdiffSref}{1.01}
\newcommand{\sgdiffScorr}{0.71}
\newcommand{\sgsameSref}{0.93}
\newcommand{\sgsameScorr}{0.64}

\newcommand{\Nmaxatoms}{200}
\newcommand{\Nnoncubic}{15{,}560}
\newcommand{\relaxDV}{2.92}
\newcommand{\relaxDS}{1.28}
\newcommand{\relaxDVPBE}{4.11}
\newcommand{\relaxDSPBE}{0.93}

\newcommand{\allrscann}{8{,}803}
\newcommand{\allrscanmed}{0.43}
\newcommand{\allrscanPBEmed}{0.44}
\newcommand{\tailrscanmed}{2.99}
\newcommand{\tailrscanPBEmed}{4.09}

\newcommand{\hpVFin}{1.24}

\newcommand{\htVFin}{1.21}

\newcommand{\TsymExact}{9{,}132}

\newcommand{\TgcBrelax}{1.16}
\newcommand{\TgcBmodel}{0.75}

\newcommand{\TgcPBcell}{1.04}

\newcommand{\TgcPBmodel}{0.86}

\newcommand{\Tsprrho}{0.68}
\newcommand{\Tsprrhonc}{0.52}

\newcommand{\Tsprratio}{2.6}

\newcommand{\memIVrel}{14}

\newcommand{\memISrel}{20}

\newcommand{\memIBrel}{2}

\newcommand{\memIn}{32}
\newcommand{\memRSdftinp}{0.94}
\newcommand{\memRSdfttight}{1.20}

\newcommand{\memRSredinp}{26}
\newcommand{\memRSredtight}{30}

\newcommand{\rAMendS}{0.64}

\newcommand{\rAHeadthreeV}{0.92}
\newcommand{\rAHeadthreeS}{0.71}

\newcommand{\rABestthreeV}{0.88}
\newcommand{\rABestthreeS}{0.65}

\newcommand{\rAtB}{0.77}
\newcommand{\rAtBdir}{0.86}

\newcommand{\kapK}{30}
\newcommand{\kapBK}{30}

\newcommand{\mlipKBtece}{10}

\newcommand{\mlipKtece}{100}

\newcommand{\esV}{0.83}
\newcommand{\esS}{0.64}
\newcommand{\esB}{0.85}

\makeatletter
\def\email#1{\global\advance\emailcnt by 1\relax%
\if@corauemail%
   \ifx\corrauthemail\@empty\else\g@addto@macro\corrauthemail{;\ }\fi%
   \g@addto@macro\corrauthemail{\setcounter{footnote}{0}\textcolor{blue}{#1}}%
\else%
   \ifx\authemail\@empty\else\g@addto@macro\authemail{;\ }\fi%
   \g@addto@macro\authemail{\setcounter{footnote}{0}\textcolor{blue}{#1}}%
\fi}
\makeatother

\begin{document}
\predisplaypenalty=0
\nocite{hohenberg1964,kohn1965,pbe1996,kurth1999_errorpbe,Lejaeghere2014_errorpbe,isaacs2018,lejaeghere2016,jain2013,dudarev1998,chen2022_m3gnet,deng2023_chgnet,batatia2022_mace,batatia2024_macemp0,kellner2026,mannan2026_uniffbench,sun2015_scan,furness2020_r2scan,kingsbury2022_r2scan,balabin2009,ramakrishnan2015,hu2003,simak2025,gong2022,hussein2022,yuk2024,vandewalle1999,rocken2024,yang2024_deeprelax,yang2026_e3relax,fogarasi1992,melli2021,lemm2021_g2s,yoon2020_dogss,senior2020_alphafold,yang2020_trrosetta,hiranuma2021,jack1978,zubatyuk2025,wallace1972,grazulis2012_cod,meier1969,barrosoluque2024_omat24,grimme2010_d3,tece2026,equflashv2_ckpt,riebesell2025_matbench,matbench_leaderboard,allen1986,woinska2016,zagorac2019_icsd,fu2025_esen,grimme2011_bj,larsen2019_dim,wu2005_relaxedion,dejong2015_elastic,klipfel2023,zeni2025_mattergen,waser1963_restraints,mehl2017_aflow,field2007_clusterboot,kresse1996_prb,kresse1996_cms,paszke2019_pytorch,larsen2017_ase,ong2013_pymatgen,togo2018_spglib}

\title[Approaching experimental accuracy for crystal structures]{RECAST: Approaching experimental accuracy in theoretical calculations of crystal structures}

\author*[1,]{\fnm{Utkarsh} \sur{Singh}}\email{utkarsh.singh[at]liu.se}
\author[2]{\fnm{Pamela} \sur{Costa Carvalho}}
\author[2,3]{\fnm{Erna K.} \sur{Delczeg-Czirjak}}
\author[2,3]{\fnm{Olle} \sur{Eriksson}}
\author[1]{\fnm{Sergei I.} \sur{Simak}}

\affil[1]{\orgdiv{Department of Physics, Chemistry and Biology (IFM)}, \orgname{Linköping University}, \orgaddress{\postcode{SE-581 83}, \city{Linköping}, \country{Sweden}}}
\affil[2]{\orgdiv{Department of Physics and Astronomy}, \orgname{Uppsala University}, \orgaddress{\street{Box 524}, \postcode{SE-75120}, \city{Uppsala}, \country{Sweden}}}
\affil[3]{\orgdiv{WISE - Wallenberg Initiative Materials Science for Sustainability, Department of Physics and Astronomy}, \orgname{Uppsala University}, \orgaddress{\postcode{SE-751 20}, \city{Uppsala}, \country{Sweden}}}

\abstract{\unboldmath
Many physical properties of materials are highly sensitive to their crystal structures.
However, density-functional theory (DFT) yields systematic errors in cell volumes, shapes and atomic positions -- a trait that is inherited by machine-learned interatomic potentials (MLIPs) trained on DFT data. Here we introduce RECAST (refinement of calculated structures), an equivariant method that learns to transform calculated structures towards their experimental counterparts.
Trained on $\Nbench$ calculated–experimental structure pairs, it predicts how each interatomic distance should change. A global structural fit incorporates these corrections as an effective energy contribution balanced against crystal stiffness, concentrating deformation in softer modes. RECAST lowers the held-out mean volume error from $\eVPBE\%$ to $\eVFin\%$ and removes a third of the cell-shape and half of the bond-length error. Training it on universal-MLIP relaxations yields comparable gains, enabling their structures to be corrected towards experiment on the fly. Calculated structures can thus approach experimental accuracy without measurement.
}

\keywords{density-functional theory; crystal structure; equivariant neural networks; machine-learned
interatomic potentials; structure refinement}

\maketitle

Density-functional theory (DFT)~\cite{hohenberg1964,kohn1965} is the standard computational route to crystal structures,
material densities and phase stabilities, and to the many properties that are evaluated at a relaxed geometry, such as
elastic constants and phonons. With semi-local exchange-correlation approximations the structural error is not random.
The Perdew--Burke--Ernzerhof (PBE) generalized-gradient approximation~\cite{pbe1996} overestimates most equilibrium
volumes, with errors of a few percent whose size depends on the chemistry of the
material~\cite{kurth1999_errorpbe,Lejaeghere2014_errorpbe,isaacs2018}. Converged equilibrium structures are reproducible
between independent DFT codes~\cite{lejaeghere2016}, so the remaining difference from experiment reflects the
approximation and the way the calculation is set up, together with physical differences between the static calculated
lattice and the measured one, such as thermal expansion and zero-point motion.

Even where it is small, the error of DFT regarding bond distances and crystal structures is troublesome, because the first
result of any experimental investigation of a new material is exactly its structure and bond distances. Accuracy in calculated structural properties is simply crucial, since all other physical and chemical properties depend on where the atoms are placed in the lattice.
In addition, any DFT error regarding the structure limits the
reliability of high-throughput predictions of new phases. Databases such as the Materials Project
(MP)~\cite{jain2013} report relaxed PBE and PBE+$U$ structures (PBE with an on-site Hubbard correction for localized
$d$ and $f$ electrons~\cite{dudarev1998}) that are reused for screening, as
starting points of further calculations and as training data of universal machine-learned interatomic potentials
(MLIPs)~\cite{chen2022_m3gnet,deng2023_chgnet,batatia2022_mace,batatia2024_macemp0}. In such training, the MLIPs inherit the bias of a reference functional towards poor geometries~\cite{kellner2026}. A recent benchmark against measured minerals found that even the best universal MLIPs yield densities whose errors exceed the tolerance of practical applications, also at ambient conditions~\cite{mannan2026_uniffbench}. Improved functionals such as
the strongly constrained and appropriately normed (SCAN) meta-generalized-gradient approximation~\cite{sun2015_scan}
and its revision r$^2$SCAN~\cite{furness2020_r2scan} reduce the bias within the theory at additional
computational cost~\cite{isaacs2018,kingsbury2022_r2scan}.

\begin{figure*}[p]
    \centering
    \includegraphics[width=\linewidth]{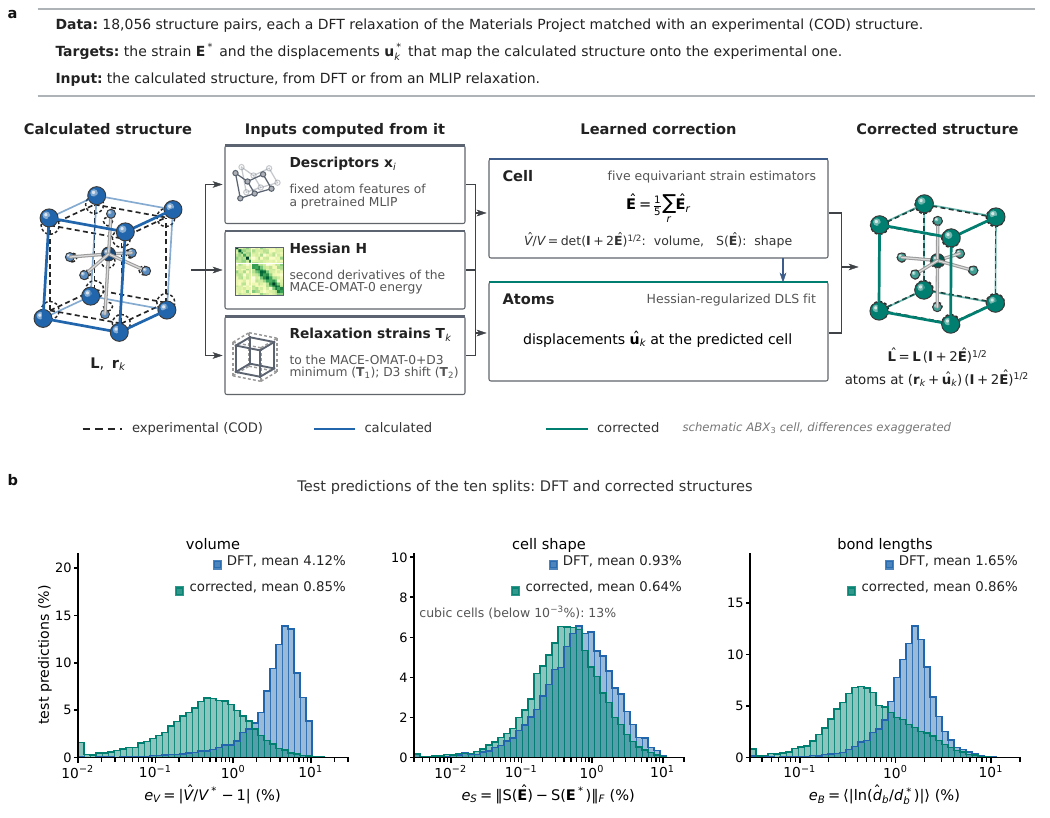}
    \caption{Correction of a calculated crystal structure towards the experimental one. \textbf{a}, A structure
    pair joins a calculated structure, here a DFT relaxation of the Materials Project, with the matching
    experimental structure of the Crystallography Open Database (COD). The
    correction predicts the strain $\bE^*$ from the calculated to the experimental cell as the mean $\hat\bE$ of five estimators
    $\hat\bE_r$, and the displacements $\bu^*_k$ of the atoms at the corrected cell by a Hessian-regularized
    distance-least-squares (DLS) fit, from inputs computed with the pretrained machine-learned interatomic potential
    (MLIP) MACE-OMAT-0, among them the strains $\bT_1$ and $\bT_2$ of two relaxations with it (Methods). In the figure, $\bL$ is the cell matrix, with the lattice vectors as rows, $\br_k$ the position of atom $k$, $\bx_i$ the descriptors of atom $i$, $\bH$ the Hessian of MACE-OMAT-0, $\hat\bu_k$ the predicted displacement of atom $k$ and $\shp(\bE)$ the shape part of a strain $\bE$, a traceless $3\times3$ matrix. \textbf{b}, Distributions of the errors $e_V$ (volume), $e_S$ (cell shape) and $e_B$ (bond
    lengths) (equation~\eqref{eq:errors} in Methods, with $V$, $\hat V$ and $V^*$ the calculated, corrected and experimental volumes, $\hat d_b$ and $d^*_b$ the predicted and experimental lengths of bond $b$, $\langle\cdot\rangle$ the mean over the bonds and $\|\cdot\|_F$ the Frobenius norm) of the DFT (blue) and corrected (teal) structures over all test
    predictions of the ten random splits, one per structure pair and split ($n=\Nrows$; $n=\NrowsB$ with matched atoms
    and bonds between atoms other than hydrogen for $e_B$). Axes are logarithmic; values outside the range are counted in the
    end bins, and shape errors below $10^{-3}\%$ (nearly all cubic cells) are given as a share. Values in the figure are means over the ten test sets.}
    \label{fig:overview}
\end{figure*}

A complementary route is to estimate the discrepancy between a computed and a reference quantity and to correct the calculated
result. For molecular systems, this was done with a simpler level of theory (such as Hartree--Fock or DFT) for the computed quantity and a higher level of theory~\cite{balabin2009,ramakrishnan2015} or experiment~\cite{hu2003} for the reference quantity.
This approach, referred to as $\Delta$-machine learning~\cite{ramakrishnan2015}, improved the formation energies. Subsequent work on solids used a similar approach, with the reference quantity being the experimental heat of formation~\cite{simak2025} (an approach termed error-corrected DFT) or its estimates~\cite{gong2022}.

For crystal volumes, a trained correction halved the
mean absolute PBE volume error of about $2{,}000$ MP structures, to $2.7\%$, and, applied as an isotropic scaling of the cell,
reduced that of the lattice constants by a factor of 3 to 5~\cite{hussein2022}; another corrected the average lattice
parameter of oxides~\cite{yuk2024}. Both correct one number, the size of the cell. Except in cubic crystals, its shape
has up to five further degrees of freedom, besides the positions of the atoms within it, which set the bond lengths and angles. An isotropic scaling leaves the shape
unchanged, and a hydrostatic pressure, used to bring a calculated volume to the measured one~\cite{vandewalle1999},
changes it only through the calculated elastic response. Interatomic potentials have been fitted to measured lattice
constants and elastic moduli of a single element~\cite{rocken2024} or simple solids~\cite{kellner2026}, and models
that predict the full cell and the atomic positions from an unrelaxed structure reproduce DFT relaxations rather than measured
structures~\cite{yang2024_deeprelax,yang2026_e3relax}. In molecules, atomic positions have been brought towards
experiment by empirical offset forces added to the ab initio energy surface~\cite{fogarasi1992} and by corrections
derived from semi-experimental equilibrium structures of related molecules~\cite{melli2021}. Predicted distances have
been turned into structures by distance geometry~\cite{lemm2021_g2s}, by relaxation in a learned harmonic force
field~\cite{yoon2020_dogss} and, in protein structure prediction, by minimizing an energy with learned
restraints~\cite{senior2020_alphafold,yang2020_trrosetta}; the predicted errors of the residue-pair distances of a
protein model guide its refinement in the same way~\cite{hiranuma2021}. Refinements of macromolecular structures
against experimental data also minimize an energy together with the misfit~\cite{jack1978}, recently with restraints
from an MLIP~\cite{zubatyuk2025}.

For a full crystal structure, the quantity to learn is the deformation that maps the calculated structure onto the experimental one: a Green--Lagrange strain~\cite{wallace1972} that adjusts the unit cell's volume and shape, together with displacements of the atoms inside it. Learning this deformation directly raises two problems that scalar properties do not. First, the correction must rotate with the crystal and preserve its symmetry, so that a cubic crystal remains cubic.
Second, because the atoms are coupled through chemical bonds, their coordinates respond collectively to any local error and cannot be corrected one at a time.

Here we introduce RECAST (refinement of calculated structures), which learns this deformation from $\Nbench$ pairs of DFT-relaxed and experimental crystal structures, taken from MP and the Crystallography Open Database (COD)~\cite{grazulis2012_cod}. Rather than predicting the 3D deformation directly, RECAST solves both problems in two steps. Its neural networks predict only rotation-invariant quantities, mainly the relative change of each interatomic distance. A physics-informed fit then reconstructs the 3D deformation that best matches these predicted distances, automatically preserving crystal symmetry. Much like distance-least-squares (DLS) refinement in crystallography~\cite{meier1969}, the fit adjusts the cell and all atomic positions simultaneously, but regularizes the solution with the crystal's stiffness (the Hessian of an MLIP). Physically, this balances the learned correction against the calculated energy landscape, allowing all atoms to move together and the crystal to yield along its softest modes. 
This fit places the atoms. The cell is corrected by the mean of five strain estimates: two from Hessian-regularized fits, two from plain DLS fits and one built from the bond directions.
All inputs are computed from the calculated structure with an MLIP trained only on the Open
Materials 2024 (OMat24) data set~\cite{barrosoluque2024_omat24}, whose structures derive from DFT relaxations rather
than from experiment. On the test sets of ten random splits, the mean errors fall from $\eVPBE\%$ to $\eVFin\%$ in
volume, from $\eSPBE\%$ to $\eSFin\%$ in cell shape and from $\eBPBE\%$ to $\eBFin\%$ in bond length
(Fig.~\ref{fig:overview}). They are at most slightly larger when whole structure prototypes or chemical systems are held out of
training. Retrained on relaxations with universal MLIPs instead of DFT, the method lowers the errors of those relaxations by similar shares.

\section*{Results}
\label{sec:results}

In the following sections we make a detailed comparison between results of RECAST and experimental results, regarding structural properties, bond distances and equilibrium volumes. Details of the methods are given in the Methods section: definitions of targets and error measures, input features and training. The corrections can be made from DFT calculations or, after retraining, from structures predicted by MLIPs.

\subsection*{Accuracy against experimental structures}

\begin{figure*}[t]
    \centering
    \includegraphics[width=\linewidth]{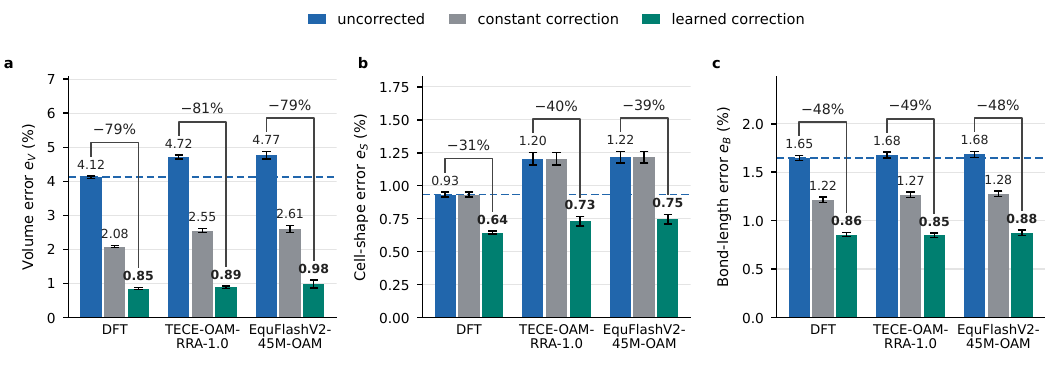}
    \caption{Accuracy of the corrected structures. Mean errors over the test sets for \textbf{a}, the volume,
    \textbf{b}, the cell shape and \textbf{c}, the bond lengths (equation~\eqref{eq:errors} in Methods), for DFT structures
    (relaxations of the Materials Project) and for relaxations of the experimental structures of the pairs with the
    machine-learned interatomic potentials (MLIPs) TECE-OAM-RRA-1.0~\cite{tece2026} and
    EquFlashV2-45M-OAM~\cite{equflashv2_ckpt}: before correction (blue), after a constant correction (grey; one
    isotropic scaling of all cells per split, which leaves the shape unchanged and scales all bond lengths with the cell) and after the learned correction, trained for
    that source (teal). Bars are means over the ten test sets ($n=10$) and error bars standard deviations across them;
    brackets give the relative change from blue to teal, and the dashed line marks the uncorrected DFT error. For DFT, the relative changes of the mean errors over the test pairs, each averaged over the splits that test it, have $95\%$ percentile
    intervals, from a bootstrap over structure prototypes, of $\randVrello$ to
    $\randVrelhi\%$ (volume), $\randSrello$ to $\randSrelhi\%$ (shape) and $\randBrello$ to $\randBrelhi\%$ (bonds).
    All errors are listed in Supplementary Table~1.}
    \label{fig:accuracy}
\end{figure*}

The correction removes most of the volume error of DFT structures and a large part of their shape and bond-length errors (Figs.~\ref{fig:overview}b and~\ref{fig:accuracy}). On the test sets of ten random splits, it lowers the mean volume, shape and bond-length errors by $\redV\%$, $\redS\%$ and $\redB\%$, respectively. The medians, below which half
of the structures lie, fall from $\eVPBEmed\%$ to $\eVFinmed\%$, from $\eSPBEmed\%$ to $\eSFinmed\%$ and from $\eBPBEmed\%$ to $\eBFinmed\%$, respectively; they are lower than the means because a few structures have large errors. The internal geometry improves as well: the root-mean-square displacement of the atoms decreases from $\eDPBE$~\AA\
to $\eDFin$~\AA\ and the bond-angle error from $\eTPBE^\circ$ to $\eTFin^\circ$. All error measures (defined in the Methods section) that describe the deviation of the corrected structure from the measured one are lower than for
DFT in each of the ten splits (Supplementary Section~1 and Supplementary
Tables~1 and~2).

Simpler corrections fall short of this. As illustrated in Fig.~\ref{fig:accuracy}a, one isotropic scaling of all cells, fitted to the training structures 
(constant correction),
lowers the volume error only to $\constVdft\%$.
A hydrostatic pressure that brings each cell to its experimental volume through the elastic response of the pretrained MLIP MACE-OMAT-0~\cite{batatia2022_mace,batatia2024_macemp0} raises the shape error (Supplementary Table~3). Relaxations with this potential and the D3 dispersion correction~\cite{grimme2010_d3} give cells with a smaller volume error than DFT ($\relaxDV$ against $\relaxDVPBE\%$) but a
larger shape error ($\relaxDS$ against $\relaxDSPBE\%$) on the same test
predictions.

The experimental
structures of the pairs were relaxed with TECE-OAM-RRA-1.0~\cite{tece2026} and EquFlashV2-45M-OAM~\cite{equflashv2_ckpt}, two of the
best-performing models on Matbench Discovery~\cite{riebesell2025_matbench,matbench_leaderboard}. Retrained on these relaxations, the correction lowers their errors (Fig.~\ref{fig:accuracy}) by
$\redeqfV$--$\redteceV\%$ (volume), $\redeqfS$--$\redteceS\%$ (shape) and $\redeqfB$--$\redteceB\%$ (bonds).
Before correction, these relaxations (force criterion $0.01$~eV\,\AA$^{-1}$) have larger volume and
shape errors than the DFT structures. Both potentials were pretrained on the relaxation trajectories of MP, which hold a frame matching the experimental structure of about half of the test pairs. On the $\expteceOutn$ pairs whose experimental structure they do not hold, the correction lowers the errors by at least the same share, so its gains do not rest on the potentials having seen these structures. The corrected errors are larger on these pairs, as they are for DFT (Supplementary
Table~4).

\subsection*{Missing-energy distance fit}

\begin{figure*}[p]
    \centering
    \includegraphics[width=\linewidth]{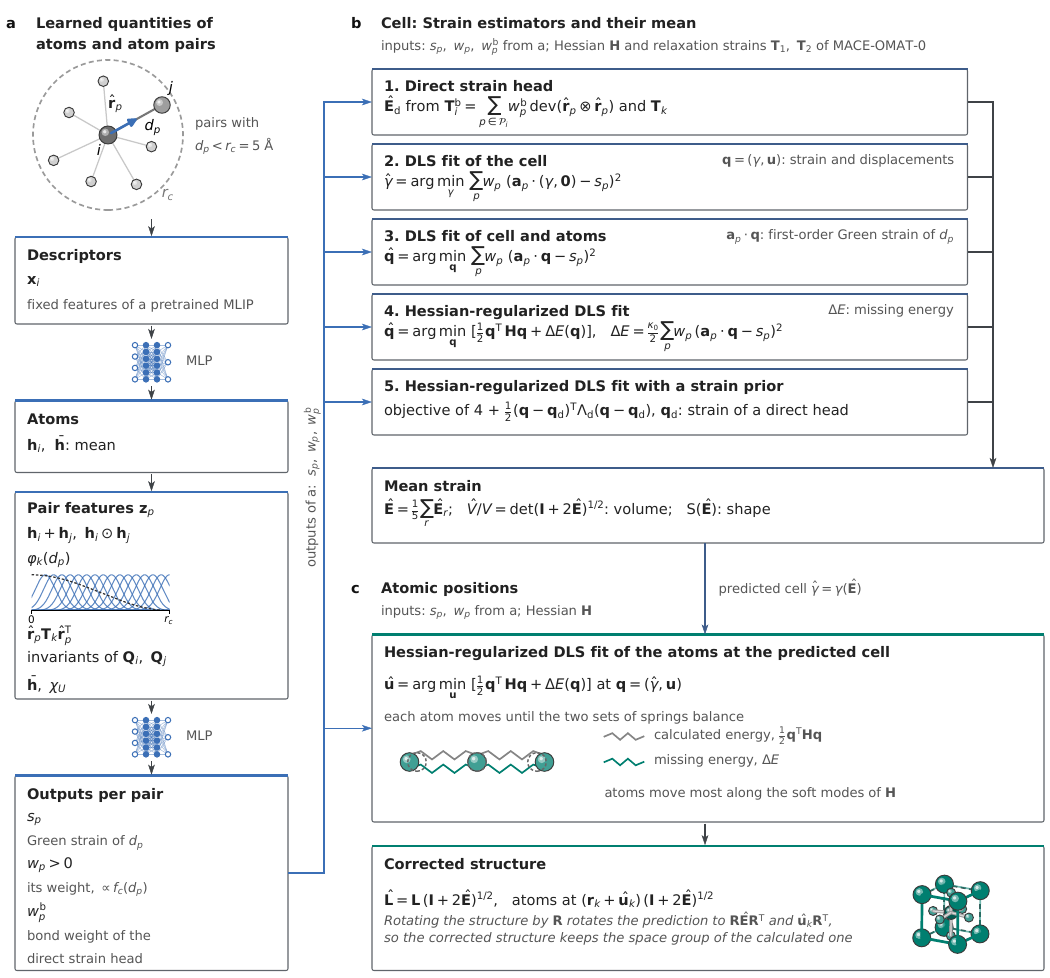}
\caption{Architecture of the correction. Networks (\textbf{a}) predict how much the length of each pair of neighbouring atoms should change. Their outputs (blue arrows) feed five strain estimates, whose mean gives the cell (\textbf{b}), and a fit that places the atoms (\textbf{c}). 
This fit uses the predicted cell (grey-blue arrow). \textbf{a}, A multilayer perceptron (MLP) maps the descriptors $\bx_i$ of atom $i$,
from the pretrained MLIP MACE-OMAT-0, to $\bh_i$; a second MLP maps the invariant features $\mathbf z_p$ of each pair $p$ within the cutoff $r_c$ to the predicted change $s_p$ of its length, its weight $w_p$ and the bond weight $w^{\rm b}_p$. 
Each estimator and the atom fit have their own such networks. 
\textbf{b}, The five estimates $\hat\bE_r$ and their mean $\hat\bE$: 1, from the direct strain head, built from bond-orientation tensors $\bT^
{\rm b}_i$; 2 and 3, from distance-least-squares (DLS) fits; 4 and 5, from Hessian-regularized DLS fits, in which the Hessian $\bH$ of MACE-OMAT-0 models the calculated energy and restraints on the pair lengths model the missing energy $\Delta E$.
\textbf{c}, The atoms at the predicted cell: grey springs, the calculated energy, at rest in the calculated structure (dashed circles); teal springs, the missing energy, with rest lengths changed by $s_p$. 
In the illustration, $\hatr_p$ and $d_p$ are the direction and length of pair $p$; $\varphi_k$, Gaussian radial basis functions; $f_c$, cutoff function; $\bar\bh$, mean of the $\bh_i$; $\mathbf Q_i$, learned atomic tensors; $\chi_U$, DFT+$U$ flag; $\mathcal P_i$, pairs of atom $i$; $\dev$, traceless part; $\bT_1$ and $\bT_2$, relaxation strains; $\hat\bE_{\rm d}$, strain of a direct head; $\kappa_0$, restraint strength, set for each structure in estimator~5, whose prior has weights $\bm\Lambda_{\rm d}$; other notation as in Methods.}

    \label{fig:arch}
\end{figure*}

\begin{figure*}[t]
    \centering
    \includegraphics[width=\linewidth]{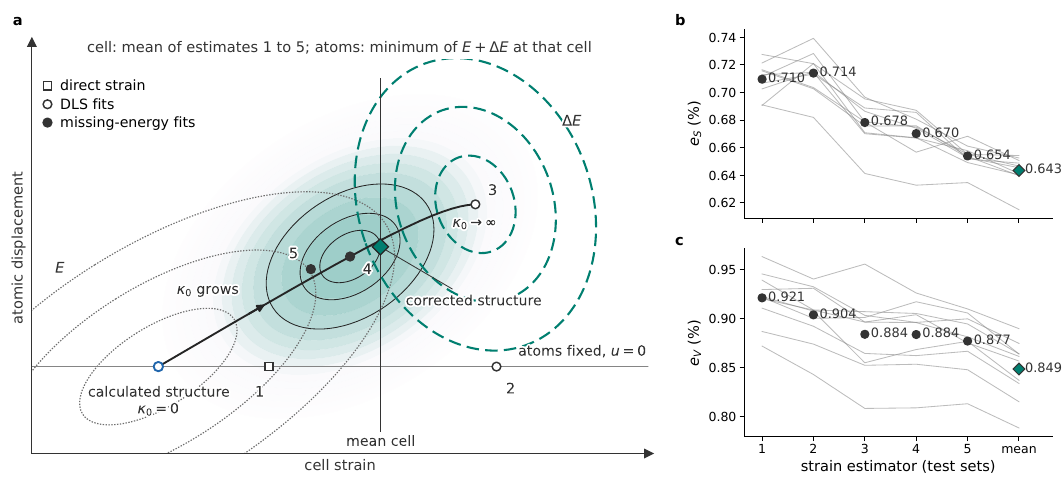}
    \caption{The five strain estimators. \textbf{a}, Schematic, in arbitrary units, with one calculated energy and
    one missing energy for all estimates. Grey dotted ellipses, the quadratic model $E$ of the calculated energy; teal
    dashed ellipses, the missing energy $\Delta E$ (equation~\eqref{eq:restraints} in Methods); black ellipses and teal shading,
    $E+\Delta E$ at one restraint strength $\kappa_0$. The five estimates, marked by kind as in the key: 1, the direct strain; 2 and 3,
    the minimum of $\Delta E$ with the atoms fixed ($\bu=\mathbf 0$) or free; 4, the minimum of $E+\Delta E$; 5, the
    same with a prior towards the direct strain. Black curve, the minima of $E+\Delta E$ for $\kappa_0$ from $0$ (the calculated structure) to
    $\infty$ (estimate 3); the arrow points to larger $\kappa_0$. The corrected cell is the mean of the five (vertical line), and the atoms take the lowest point of
    $E+\Delta E$ on that line (filled diamond). In the trained correction, each estimator and the atom fit have their own networks.
    \textbf{b},\textbf{c}, Shape error $e_S$ (\textbf{b}) and volume error $e_V$ (\textbf{c}) on the test sets of each
    estimator alone, with three initializations, and of the mean of the five (diamond; fifteen networks). Grey
    lines, the ten splits; markers, means over the ten test sets (Supplementary Table~5).}
    \label{fig:hessfit}
\end{figure*}

The correction works through distances. It predicts a relative length change for each atom pair within a cutoff of
$5$~\AA\ (Fig.~\ref{fig:arch}a), and least-squares fits turn these changes into a deformation of the cell and the atoms (Fig.~\ref{fig:arch}b,c). A
plain DLS fit weights soft and stiff deformations alike, whereas a real crystal yields most along its soft modes. Adding a
quadratic model of the calculated energy, built from the Hessian of MACE-OMAT-0, penalizes each deformation by its
stiffness and makes the predicted changes a missing energy balanced against the calculated one
(Fig.~\ref{fig:hessfit}a). In least-squares terms, this missing-energy distance fit is a Hessian-regularized DLS fit.
The cell correction is the mean of five estimators (Methods; numbered as in Fig.~\ref{fig:hessfit}a). The first builds the strain from the bond directions
around each atom. The second and third are plain DLS fits, with the atoms kept at their fractional coordinates or free to move. The fourth and fifth are
missing-energy fits, which differ in whether a direct strain prediction enters as a prior. A separate
missing-energy fit places the atoms at the corrected cell, by minimizing the sum of the calculated and missing energies over the atomic positions (filled diamond in Fig.~\ref{fig:hessfit}a).

The inputs are per-atom descriptors of MACE-OMAT-0,
its Hessian and two strains from relaxations with it: $\bT_1$, from the calculated structure to its minimum with the D3
dispersion correction, and $\bT_2$, the change of this strain when D3 is added. A variant whose cell estimators take their descriptors from the potential eSEN-30M-OMat instead (Methods) gives errors of $\esV\%$, $\esS\%$ and $\esB\%$ in volume,
shape and bond length, against $\eVFin\%$, $\eSFin\%$ and $\eBFin\%$ for the correction reported here.
The inputs, the estimators and the hyperparameters were chosen by cross-validation on the $\Ndev$ pairs that are in none of
the test sets (Methods and Supplementary Section~2); there, the two descriptor sources gave nearly equal errors, so the correction uses one potential, MACE-OMAT-0, for its descriptors and its Hessian.

The test sets, which played no part in these choices, show how the estimators compare
(Fig.~\ref{fig:hessfit}b,c).  Alone, they reach shape errors from $\rAHeadthreeS\%$ (direct strain) to $\rABestthreeS\%$ (missing-energy fit with the prior) and volume errors of $\rABestthreeV$--$\rAHeadthreeV\%$; their mean reaches $\rAMendS\%$ and $\eVFin\%$ (Supplementary Table~5).

All parts of the correction are equivariant, so the corrected structure keeps the symmetry of the calculated one by
construction (Methods). The $\TsymExact$ corrected test structures with an unambiguous space group retain it at a
tolerance of $10^{-3}$~\AA\ (Supplementary Section~3).

\subsection*{Cell shape}

\begin{figure*}[t]
    \centering
    \includegraphics[width=\linewidth]{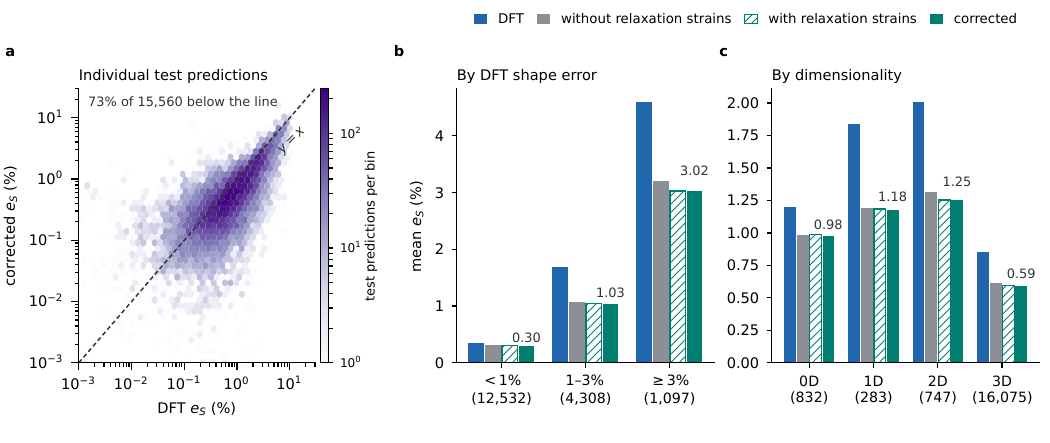}
    \caption{Cell-shape errors of the test structures. Shape error $e_S$ (equation~\eqref{eq:errors} in Methods) of all test
    predictions of the ten splits, one per structure pair and split. \textbf{a}, Corrected against DFT shape error for
    the $\Nnoncubic$ test predictions where the latter exceeds $10^{-3}\%$ (nearly all others are cubic cells,
    whose shape is fixed by symmetry); colour, number of test predictions per hexagonal bin on a logarithmic scale; below the
    dashed diagonal the error is reduced. \textbf{b}, Mean shape error in three bands of the DFT value, for DFT
    (blue), for the mean of the five strain estimators with one initialization each, without (grey) and with (hatched)
    the relaxation strains $\bT_1$ and $\bT_2$, and for the full correction, with three initializations per estimator (teal).
    \textbf{c}, The same by the dimensionality of the bonded atomic network of the DFT
    structure: molecular (0D), chains (1D), layers (2D) and bonded in three dimensions (3D). Numbers of test
    predictions in parentheses.}
    \label{fig:cell}
\end{figure*}

The shape correction removes the largest error where DFT is least accurate (Fig.~\ref{fig:cell}a,b). Of the
$\Nnoncubic$ test predictions, pooled over the ten splits, whose DFT shape error exceeds $10^{-3}\%$ (shown in Fig.~\ref{fig:cell}a, nearly all of
them non-cubic), the correction improves
$\impS\%$ (the points below the diagonal). Where that error exceeds $3\%$, it brings the mean from $\tailPBE$ to $\tailFin\%$. Below $1\%$, where $\bandshare\%$ of the
test predictions lie, the gain is small, and the error rises for $\bandworse\%$ of those above $10^{-3}\%$ in this
band.

The relaxation strains add little to the correction. Without them, the five estimators, with one initialization
each, already reach $\eSnoD\%$ in shape and $\eVnoD\%$ in volume, against $\eSone$ and $\eVone\%$ with them. The strains help most in layered (2D) materials, in which semi-local DFT
describes the interlayer binding poorly (Fig.~\ref{fig:cell}c and Supplementary
Table~6).

Large shape errors are rare but carry much of the total: only $\tailpopshare\%$ of pairs have a DFT shape error above
$3\%$, yet they account for $\tailerrshare\%$ of the summed error, and their corrected errors remain the largest. Errors this large
could stem from the experimental reference, and $\tailid\%$ of such structure pairs have atomic displacements above $0.3$~\AA\ or a changed space group (Methods), more than twice the share among all pairs.
Where the phase has a second experimental report ($\tailcodn$ of these pairs), however, the closest report agrees with
the first to a median of $\tailcod\%$ in shape, against $\tailcodpbe\%$ for DFT. Where this can be checked, the error lies
with the calculation.
Relaxations with r$^2$SCAN shrink these errors without changing their direction: where MP holds such relaxations
($\tailrscann$ of these pairs), the median shape error falls from $\tailrscanPBEmed$ to $\tailrscanmed\%$, and the
median cosine between their traceless strain errors and those of DFT is $\tailrscan$. Over all $\allrscann$ pairs with
r$^2$SCAN relaxations, the median shape error is close to that of DFT ($\allrscanmed$ against
$\allrscanPBEmed\%$). The large errors are concentrated in structures that are not bonded in three dimensions
($\tailnonthreeD\%$ of them, against $\allnonthreeD\%$ of all pairs), in hydrogen-containing materials and in elastically
soft ones, consistent with the dispersion and hydrogen bonding that semi-local functionals miss (Supplementary
Section~1).

Two kinds of structure retain larger errors, by construction. The correction keeps the symmetry of the DFT structure,
which differs from that of the experimental one for $\sgdiffshare\%$ of the pairs; their shape error still falls, from
$\sgdiffSref$ to $\sgdiffScorr\%$, against $\sgsameSref$ to $\sgsameScorr\%$ for the others. It also has no pressure or
temperature input, so high-pressure and high- or low-temperature phases have larger volume errors, $\hpVFin$ and
$\htVFin\%$ against $\eVFin\%$ overall (Supplementary Table~6).

A correction should also signal when it is unreliable. The spread $s_S$ of the five cell
estimates, the root-mean-square deviation of their traceless parts from the mean, does so: its rank correlation with the
corrected shape error is $\Tsprrho$ ($\Tsprrhonc$ above a DFT shape error of $10^{-3}\%$), and in the top tenth of $s_S$
that error averages $\Tsprratio$ times its overall mean.

\subsection*{Atomic positions}

Most of the bond-length gain comes from the cell, and the rest from the displacements. At the corrected cell, the
bond-length error is $\TgcPBcell\%$ with the atoms at their DFT fractional coordinates, and the predicted displacements
lower it to $\TgcPBmodel\%$ (Supplementary Section~1). No correction of the cell alone reaches this: an
isotropic scaling to the experimental volume, the limit of an ideal volume-only correction~\cite{hussein2022}, leaves
$\eBIso\%$, and even the experimental cell, without displacements, gives $\eBMeas\%$ (Supplementary Table~7).

Although the correction uses the Hessian of MACE-OMAT-0, the learned displacements are not those of a relaxation with this potential. At the experimental cell, they lower the bond-length error to $\TgcBmodel\%$, whereas relaxing the atoms raises it to $\TgcBrelax\%$. The Hessian contributes to this accuracy: an atom network whose predicted bond tensions move the atoms through the Hessian gives $\rAtB\%$ there, and the same network predicting the displacements directly is worse in all ten splits, with $\rAtBdir\%$ (Supplementary Section~1).

Bonds to hydrogen improve as well. X-ray diffraction locates hydrogen by its electron density, which is shifted into the
bond~\cite{allen1986,woinska2016}, so these bonds are evaluated on the $\XHn$ neutron structures among the test pairs.
Their error falls from $\XHPBE$ to $\XHFin\%$, below the $\constXHdft\%$ of a constant scaling of the cell, although the displacements are trained with no weight on hydrogen (Supplementary Tables~1
and~7).

\subsection*{Unseen structure prototypes and chemical systems}

To test whether the correction transfers to structures unlike those it has seen, the same model was retrained in
five-fold cross-validation that holds out whole groups of pairs: in one
family of folds the $7{,}432$ structure prototypes (anonymized formula, space group and Wyckoff letters; Methods), in the other the $11{,}626$ chemical systems (sets of elements), each held
out once. Each pair is thus predicted by a model that saw no pair of its prototype (or chemical
system). The largest prototype is the ThCr$_2$Si$_2$ type, with $252$ pairs, and the largest chemical system is silica (O--Si), with $44$.
The corrected errors are at most slightly larger than on the random splits: $\gPVcorr\%$ (volume), $\gPScorr\%$ (shape) and $\gPBcorr\%$ (bonds) with prototypes held out and $\gCVcorr\%$, $\gCScorr\%$ and $\gCBcorr\%$ with chemical systems held out, against $\eVFin$, $\eSFin$ and $\eBFin\%$. All error measures are lower than for DFT in each fold (Supplementary Table~8).
The larger shape error with prototypes held out mainly reflects the extrapolation to unseen structure types, not the smaller training sets of these folds ($72\%$ of the pairs, against $81\%$; Supplementary Table~5). Beyond $16$ atoms, the relative gains show no trend with cell size, up to $\Nmaxatoms$ atoms (Supplementary Table~6).

\subsection*{Comparison with scatter in experimental data}

Experiment sets a scale for these errors, since two measurements of the same phase also
 differ. For 
$\shapesubn$ test pairs whose phase has a second experimental report in the COD (Supplementary Section~4), their median shape errors are $\shapesubPBE\%$ (DFT), $\shapesubFin\%$ (corrected) and $\shapesubIR\%$ (second report). For a subset of these with matched atoms and bonds without hydrogen ($\codsubn$ pairs), the median bond-length errors are $\codsubPBE\%$ (DFT), $\codsubFin\%$ (corrected) and $\codsubIR\%
$ (second report). The corrected bond lengths are
thus typically closer to the experimental structure than a second report is, whereas the corrected cell shapes are not. Part of the difference between two reports is physical, since they can come from different samples or temperatures.

\section*{Discussion}
\label{sec:discussion}

The correction shifts the volume, the cell shape and the atomic positions of a calculated structure towards experiment, and
its construction is tied neither to the computational method that produced the structure nor to a specific class of materials
(Fig.~\ref{fig:arch}). Its predicted distance
changes act as a missing energy, balanced against the stiffness that the Hessian of an MLIP gives the calculated structure.

A corrected structure estimates the experimental one, which no static calculation targets. A measured structure also
reflects thermal expansion, zero-point motion, differences between samples and the uncertainties that originate from experimental analysis. An improved functional does not
close this gap on its own. In MP, r$^2$SCAN relaxations shrink the largest DFT shape errors by about a quarter but keep
their direction and leave the typical error unchanged, whereas the correction lowers both.

The correction applies to the kind of data it is trained on. Relaxations with another code, functional or interatomic potential require new structure pairs and retraining, but not a new method. So do experimental structures from another collection such as the Inorganic Crystal Structure Database~\cite{zagorac2019_icsd}, whose additional data points could improve the corrections.

The missing energy is learned from geometries and acts, with the Hessian, as a quadratic model at the calculated structure,
and the elastic moduli of that Hessian agree closely with those of MP (Methods). 
The gain of the correction does not rest on MP relaxations retaining part of their starting geometry (Methods).
The accuracy was measured on pairs whose DFT volume and shape are within $10\%$ of experiment, and where DFT is already accurate the correction can add error; the spread of the five cell estimates marks the
predictions that one should trust least. The corrected shape and volume errors remain larger for molecular, chain and
layered crystals, $\allnonthreeD\%$ of the pairs, than for 3D-bonded ones (Fig.~\ref{fig:cell}c and Supplementary Table~6). The errors still fall as training pairs are added (Supplementary Table~5); more experimental structures of such materials may lower them further.

The results presented here show that much of the error of calculated crystal structures and bond distances can be predicted and removed. The correction eliminates nearly four-fifths of the volume error of DFT structures and about a third to a
half of their shape and bond-length errors. It keeps most of this gain for structure prototypes and chemical systems
absent from training, gives similar gains for two universal MLIPs when retrained, 
and preserves the symmetry of all crystals tested here. After correction, the mean volume errors of the DFT structures and of both MLIP relaxations are below $1\%$.
Calculated structures underlie most computed properties and fill the databases used to screen materials; since the
correction needs only MLIP calculations and no measured input, it can be applied wherever structures are computed in
bulk with the methods it was trained for. Properties computed at corrected structures, such as elastic constants,
phonons, band edges or magnetic moments, can then show how much of their error stems from the geometry. Fitting the
missing energy to energies and forces as well as to geometries could correct the energy surface itself. The trained networks and a web application that corrects calculated structures are openly available, and the code will be released upon acceptance (Code availability).

\FloatBarrier
\section*{Methods}
\label{sec:methods}

We predict a deformation, a strain of the cell and displacements of the atoms, that carries a calculated structure
onto its experimental counterpart. Because it must transform with the crystal under rotations and reflections,
multilayer perceptrons (MLPs) predict only invariants, mainly a relative change $s_p$ of the length of each
interatomic pair $p$ and a weight $w_p$, and equivariant fits turn them into the deformation (Fig.~\ref{fig:arch}). The
main one,
the missing-energy distance fit, is a distance-least-squares (DLS) fit regularized by the Hessian of an MLIP
(MACE-OMAT-0; Supplementary Table~9 compares it with eSEN-30M-OMat).

\subsection*{Deformation and targets}
\label{sec:m_labels}
\emph{Notation.} A cell matrix $\bL$ has the lattice vectors (\AA) as rows; atom $k$ of the $n$ atoms in the cell has
fractional coordinates $\mathbf f_k$ (a row) and position $\br_k=\mathbf f_k\bL$. The metric tensor is
$\bG=\bL\bL^{\mathsf T}$ and the volume $V=|\det\bL|$. For a $3\times3$ matrix $\mathbf X$,
$\dev\mathbf X=\mathbf X-\tfrac13(\tr\mathbf X)\I$ is its traceless part, with $\tr$ the trace and $\I$ the identity,
and $\|\mathbf X\|_F$ its Frobenius norm; $\hatr\otimes\hatr=\hatr^{\mathsf T}\hatr$ for a unit row vector $\hatr$, and
functions of symmetric matrices act on their eigenvalues. A rotation or reflection
$\mathbf R\in O(3)$ maps positions as $\br\to\br\mathbf R^{\mathsf T}$ and strains as
$\bE\to\mathbf R\bE\mathbf R^{\mathsf T}$.

\emph{Deformation.} A symmetric strain $\bE$, with $\I+2\bE$ positive definite, and displacements $\bu=(\bu_1,\dots,\bu_n)$ deform a calculated structure
into the structure with cell and positions
\begin{equation}
\bL(\bE)=\bL\,(\I+2\bE)^{1/2},\qquad \br_k(\bE,\bu)=(\br_k+\bu_k)\,(\I+2\bE)^{1/2}:
\label{eq:pred}
\end{equation}
the atoms are displaced within the calculated cell, and the cell and its atoms are then strained together. The
volume becomes $VJ(\bE)$, with $J(\bE)=\det(\I+2\bE)^{1/2}$, at any size of strain, so that $\ln J$ is exactly the
volume part of the deformation and $\shp(\bE)=\tfrac12\ln\!\big[J^{-2/3}(\I+2\bE)\big]$ its shape part. A predicted
deformation $(\hat\bE,\hat\bu)$ gives the corrected structure, with cell $\hat\bL=\bL(\hat\bE)$, volume $\hat V$ and
positions $\br_k(\hat\bE,\hat\bu)$.

\emph{Targets.} The targets $(\bE^*,\bu^*)$ are the deformation that carries the calculated structure onto the
experimental structure of its pair, with cell $\bL^*$ and fractional coordinates $\mathbf f^*_k$ in the same lattice
basis and atom order (\emph{Structure pairs}), up to an orthogonal transformation and a translation. With $\bG^*=\bL^*\bL^{*\mathsf T}$
and $V^*=|\det\bL^*|$, the strain target is the Green--Lagrange strain
\begin{equation}
\bE^*=\tfrac12\big(\bL^{-1}\bG^*\bL^{-\mathsf T}-\I\big),\qquad \bL^*=\bL\,(\I+2\bE^*)^{1/2}\mathbf R_0,\qquad
J(\bE^*)=V^*/V,
\label{eq:label}
\end{equation}
a symmetric tensor in the Cartesian frame of the calculated cell; the orthogonal factor $\mathbf R_0$ of the polar
decomposition of $\bL^{-1}\bL^*$ is discarded. With the fractional coordinates of the
calculated structure unwrapped to the periodic image nearest the experimental ones, the displacement target is
\begin{equation}
\bu^*_k=(\mathbf f^*_k-\mathbf f_k)\bL-\frac1n\sum_{l=1}^n(\mathbf f^*_l-\mathbf f_l)\bL ,
\label{eq:ulabel}
\end{equation}
whose mean over the atoms is zero exactly.

\emph{Coordinates.} The estimators work in the coordinates $\bq=(\gam,\bu)\in\mathbb R^{6+3n}$ of the calculated
structure: the engineering Voigt strain
$\gam(\bE)=(E_{11},E_{22},E_{33},2E_{23},2E_{13},2E_{12})$,
read as the Green--Lagrange strain of equation~\eqref{eq:pred}, and the $3n$ Cartesian displacements (\AA); with
$\mathbf G_F=\mathrm{diag}(1,1,1,\tfrac12,\tfrac12,\tfrac12)$, $\gam^{\mathsf T}\mathbf G_F\gam=\|\bE\|_F^2$. The
linearizations below set up the solves and the bond term of the atom loss, whereas the cell loss and all errors use the
exact strains and lengths (Supplementary Section~4).

\subsection*{Structure pairs}
\label{sec:m_data}
Each structure pair joins the PBE or PBE+$U$ relaxation of a material in MP (database version 2026.09.28), of the run type that MP's mixing rule assigns to its composition, with a matching experimental structure in the COD, within
$10\%$ of experiment in volume per atom and in shape, $\|\shp(\bE^*)\|_F$
(Supplementary Section~5 and Supplementary Table~10). For an MLIP, the calculated structure is its relaxation of the
experimental one. Pairs with displacements above $0.3$~\AA\ or a changed space group are kept; leaving them out changes the mean test errors by about $0.01$ percentage points at most (Supplementary Section~4).

\subsection*{Error measures}
\label{sec:m_errors}
With $\hat d_b$ and $d^*_b$ the predicted and experimental lengths of bond $b$ of the set $\mathcal B$ of first-shell
bonds between atoms other than hydrogen (Supplementary Section~4), $n'$ the number of these atoms ($k\notin\mathrm H$),
$\bar{\mathbf t}$ the mean of $\hat\bu_k-\bu^*_k$ over them, the metric tensors
$\hat\bG=\bL(\I+2\hat\bE)\bL^{\mathsf T}$ and $\bG^*=\bL(\I+2\bE^*)\bL^{\mathsf T}$, and the angles
$\phi_{jk}(\bG)=\arccos\big(G_{jk}/\sqrt{G_{jj}G_{kk}}\big)$ between lattice vectors $j$ and $k$, with $G_{jk}$ the
elements of $\bG$, the errors of a predicted structure are
\begin{align}
e_V&=\big|\hat V/V^*-1\big|, & e_S&=\|\shp(\hat\bE)-\shp(\bE^*)\|_F,\nonumber\\
e_L&=\tfrac13\textstyle\sum_{i=1}^3\big|(\hat G_{ii}/G^*_{ii})^{1/2}-1\big|, &
e_A&=\tfrac13\textstyle\sum_{j<k}\big|\phi_{jk}(\hat\bG)-\phi_{jk}(\bG^*)\big|,\nonumber\\
e_B&=\frac1{|\mathcal B|}\sum_{b\in\mathcal B}\big|\ln(\hat d_b/d^*_b)\big|, &
e_D&=\Big(\frac1{n'}\sum_{k\notin\mathrm H}\|\hat\bu_k-\bu^*_k-\bar{\mathbf t}\|^2\Big)^{1/2},
\label{eq:errors}
\end{align}
where $e_V$ is the relative error of the volume, $e_S$ the error of the cell shape (the norm of the difference of
the shape parts of the strains), $e_L$ the mean relative error of the three lattice-vector lengths, $e_A$ the mean
absolute error of the three angles between lattice vectors, $e_B$ the mean absolute log ratio of the bond lengths and
$e_D$ the root-mean-square error of the positions of the atoms other than hydrogen after removal of their mean shift;
$e_\theta$ is the mean absolute error of the angles between bonds of $\mathcal B$ at each atom. $e_V$, $e_S$, $e_L$
and $e_B$ are given in $\%$, $e_A$ and $e_\theta$ in degrees and $e_D$ in \AA; for the calculated structure
$\hat\bE=\mathbf 0$ and $\hat\bu=\mathbf 0$, and $e_{XH}$ is $e_B$ for the bonds to hydrogen in the $\Nneutron$ neutron
structures, $\XHn$ of them among the test pairs.

\subsection*{Input features}
\label{sec:m_inputs}
All inputs are computed from the calculated structure (Supplementary Section~6). The descriptors $\bx_i$ of
atom $i$ are fixed features of potentials pretrained on the Open Materials 2024 (OMat24) data set~\cite{barrosoluque2024_omat24}
alone: the invariant node features of MACE-OMAT-0~\cite{batatia2022_mace,batatia2024_macemp0},
$256$ per atom, and, for a variant of the cell estimators, the rotation-invariant ($l=0$) channels $\bm\xi_i$ of the ten
layers of eSEN-30M-OMat (equivariant smooth energy network)~\cite{fu2025_esen}, $1{,}280$ per atom. Because this potential is
equivariant under rotations but not under reflections, these channels change under inversion; the variant uses
their average over the structure $\mathcal S$ and its inversion image $\mathcal I\mathcal S$,
\begin{equation}
\bx_i=\tfrac12\big[\bm\xi_i(\mathcal S)+\bm\xi_i(\mathcal I\mathcal S)\big],
\label{eq:desc}
\end{equation}
which is invariant under all rotations and reflections. Two relaxation strains with MACE-OMAT-0 enter as well, $\bT_1$ to its minimum with the D3 dispersion
correction~\cite{grimme2010_d3,grimme2011_bj} and $\bT_2$ the change of the relaxation strain when D3 is added, with masks
$\mu_k$ that are $0$, with $\bT_k=\mathbf 0$, where a relaxation did not converge or changed the space group; a flag
$\chi_U$ marks DFT+$U$ calculations. Both relaxations are run for each calculated structure, whatever the
dimensionality of its bonded network~\cite{larsen2019_dim}, since the DFT and MLIP calculations lack dispersion and D3 typically shifts the relaxed shape towards the measured one in each class; in 3D-bonded structures, however, it mainly contracts the cell evenly (Supplementary Section~6). The strain $\bT_2$ enters as an input, whose weight is learned, not as a direct
correction.

The Hessian $\tilde\bH$ of the MACE-OMAT-0 energy $E_{\rm M}(\bq)$, without D3, of the calculated structure
deformed by $\bq$
(equation~\eqref{eq:pred}) at $\bq=\mathbf 0$ is assembled from the clamped-ion elastic tensor $\mathbf C^0$, the
internal-strain coupling $\tilde\bH_{u\gamma}$~\cite{wu2005_relaxedion} and the force constants $\bm\Phi$, exact second
derivatives of $E_{\rm M}$ by automatic differentiation in double precision (Supplementary Section~6). Unstable
and zero modes are removed in the scaled coordinates $\mathbf S^{-1}\bq$, with
$\mathbf S=\mathrm{diag}\big(V^{-1/3}\mathbf G_F^{-1/2},\,\I_{3n}\big)$, so that all blocks share the unit
eV\,\AA$^{-2}$. With the eigenpairs $(\lambda_k,\mathbf v_k)$ of $\mathbf S\tilde\bH\mathbf S$, the Hessian used below is
\begin{equation}
\bH=\mathbf S^{-1}\Big(\sum_{\lambda_k>\eta}\lambda_k\mathbf v_k\mathbf v_k^{\mathsf T}\Big)\mathbf S^{-1},
\label{eq:Hplus}
\end{equation}
which keeps the modes with $\lambda_k>\eta=10^{-3}$~eV\,\AA$^{-2}$ and has zero stiffness along the removed ones;
$\bH^{+}$ inverts it on the retained modes. Its relaxed-ion elastic moduli agree with those of
MP~\cite{dejong2015_elastic} (Supplementary Section~6).

\subsection*{Atom and pair networks}
\label{sec:m_cell}
Each estimator has its own MLPs on invariants (Fig.~\ref{fig:arch}a), with $0.05$ to
$1.31$ million parameters, which embed the descriptors as $\bh_i=\mathrm{MLP}(\bx_i)$, with their mean $\bar\bh$ over the atoms, and act on
all pairs $p=(i,j,\mathbf m)$ within $r_c=5$~\AA, with $\br_p=\br_j+\mathbf m\bL-\br_i$ for the lattice translation
$\mathbf m$, $d_p=\|\br_p\|$, $\hatr_p=\br_p/d_p$ and $\mathcal P_i$ the pairs of atom $i$. From invariant features of
the pair (Supplementary Section~7), a pair network gives outputs $(o_1,o_2)$ that set the predicted
Green strain of $d_p$ and its weight,
\begin{equation}
s_p=10^{-2}\,o_1,\qquad w_p=f_c(d_p)\,\exp\!\big[\mathrm{clip}(o_2,-3,3)\big].
\label{eq:sw}
\end{equation}
The cutoff function is $f_c(d)=\tfrac12[\cos(\pi d/r_c)+1]$, and clip limits $o_2$ to $[-3,3]$; $s_p$ and $w_p$ are
invariant and symmetric in the two atoms, and the direct strain head uses bond weights $w^{\rm b}_p$, an MLP output times $f_c(d_p)$ of either sign.

\subsection*{Strain estimators}
\label{sec:m_estimators}
From the invariants, five estimators form the strain (Fig.~\ref{fig:arch}b). Estimator~1 gives it directly, as a sum
of tensors built from the bond directions. Estimators~2 to 5 predict how the length of each pair should change and
find the deformation that best realizes these changes. Estimators~2 and 3 use the DLS fit of crystallography~\cite{meier1969}, with the
atoms kept at their fractional coordinates (estimator~2) or free to move (estimator~3). Estimator~4 uses the Hessian-regularized DLS
fit, which balances these changes, as a missing energy, against the calculated energy so that the structure gives way along its
soft modes (Fig.~\ref{fig:hessfit}). Estimator~5 uses the same fit with a direct prediction of the strain as a
prior. Their mean is the cell correction.

\emph{Direct strain head (estimator~1).}\label{sec:m_head} It sums equivariant tensors with invariant coefficients, as do equivariant networks that update or generate crystal
lattices~\cite{klipfel2023,zeni2025_mattergen}.
With the bond-orientation tensors $\bT^{\rm b}_i=\sum_{p\in\mathcal P_i}w^{\rm b}_p\dev(\hatr_p\otimes\hatr_p)$ of the
atoms, its strain is
\begin{equation}
\begin{split}
\hat\bE_{\rm d}={}&10^{-2}\Big[\tfrac13\tau\,\I+\frac1n\sum_{i=1}^n\big(a_i\bT^{\rm b}_i
+c_i\dev(\bT^{\rm b}_i\bT^{\rm b}_i)\big)\Big]\\
&+\sum_{k=1}^2\mu_k\Big[\alpha_k\dev\bT_k+\tfrac13\beta_k(\tr\bT_k)\I\Big],
\end{split}
\label{eq:A}
\end{equation}
with $(a_i,c_i)=\mathrm{MLP}(\bh_i,\bar\bh,\chi_U)$, $\tau=\mathrm{MLP}(\bar\bh,\chi_U)$ and
$(\alpha_k,\beta_k)=\mathrm{MLP}(\bar\bh,\chi_U)$. Each term is an invariant scalar
times a tensor that maps as $\mathbf X\to\mathbf R\mathbf X\mathbf R^{\mathsf T}$, so that $\hat\bE_{\rm d}$ is
equivariant.

\emph{DLS fits (estimators~2 and 3).}\label{sec:m_dls} To first order in $\bq$,
the Green strain of the length of pair $p$ is
\begin{equation}
\mathbf a_p\cdot\bq=\hatr_p\,\bE\,\hatr_p^{\mathsf T}+\hatr_p\cdot(\bu_j-\bu_i)/d_p ,
\label{eq:pair}
\end{equation}
which defines the vector $\mathbf a_p\in\mathbb R^{6+3n}$. Each predicted change $s_p$ thus constrains one linear function
of $\bq$. The DLS fit is the weighted least-squares
solution, with its normal equations,
\begin{equation}
\begin{split}
\hat\bq&=\arg\min_{\bq}\Big[\sum_pw_p(\mathbf a_p\cdot\bq-s_p)^2+\bq^{\mathsf T}\bm\Lambda_0\bq\Big],\\
(\bA+\bm\Lambda_0)\,\hat\bq&=\mathbf b,\qquad \bA=\sum_pw_p\mathbf a_p\mathbf a_p^{\mathsf T},\qquad
\mathbf b=\sum_pw_ps_p\mathbf a_p ,
\end{split}
\label{eq:dls}
\end{equation}
where the small ridge term $\bm\Lambda_0=\mathrm{diag}(\lambda_g\mathbf G_F,\lambda_u\I_{3n})$, with weights $\lambda_g$
and $\lambda_u$ on the strain and the displacements (Supplementary Section~7), fixes the
rigid translation and any combination of coordinates that no pair constrains. Estimator~2 solves equation~\eqref{eq:dls} for
the strain alone, with $\bu=\mathbf 0$, and estimator~3 for the strain and the displacements.

\emph{Hessian-regularized DLS fit (estimator~4).}\label{sec:m_hessfit} The DLS fit treats soft and stiff deformations
alike. Let $E(\bq)$ be the energy of the calculated structure deformed by $\bq$ in the
method that produced it (DFT or an MLIP), and $\Delta E(\bq)$ the energy that this method misses, defined so that the experimental structure minimizes
$E+\Delta E$. We model it near the calculated structure by harmonic
restraints~\cite{waser1963_restraints} of strength $\kappa_0$ on the pair lengths, whose minima lie at the predicted
changes,
\begin{equation}
\Delta E(\bq)=\frac{\kappa_0}2\sum_pw_p\big(\mathbf a_p\cdot\bq-s_p\big)^2 ,
\label{eq:restraints}
\end{equation}
so that, to first order in $\bq$, pair $p$ acts as a spring on the Green strain of $d_p$, of stiffness $\kappa_0w_p$, whose rest length
is changed by the factor $(1+2s_p)^{1/2}$ (for $s_p>-\tfrac12$); at $\bq=\mathbf 0$, $\Delta E$ exerts the generalized force
$-\nabla\Delta E(\mathbf 0)=\kappa_0\mathbf b$, a stress on the cell and forces on the atoms, akin to the offset forces
that correct computed molecular geometries~\cite{fogarasi1992}. Because the calculated structure is a minimum of $E$, up
to the convergence of its relaxation,
$E(\bq)-E(\mathbf 0)=\tfrac12\bq^{\mathsf T}\bH_E\bq+O(\|\bq\|^3)$, with $\bH_E$ the Hessian of the method. We replace
$\bH_E$ by $\bH$, that of MACE-OMAT-0 (equation~\eqref{eq:Hplus}), with the projector $\mathbf P$ that removes the three
rigid translations; the gradient of this potential, which is not that of the method, is not used. The estimate minimizes the resulting quadratic model of $E+\Delta E$, with the ridge term of
equation~\eqref{eq:dls}:
\begin{equation}
\begin{split}
\hat\bq&=\arg\min_{\bq}\Big[\tfrac12\bq^{\mathsf T}\mathbf P\bH\mathbf P\bq+\Delta E(\bq)
+\tfrac{\kappa_0}2\bq^{\mathsf T}\bm\Lambda_0\bq\Big],\\
\mathbf M\hat\bq&=\kappa_0\mathbf b,\qquad \mathbf M=\mathbf P\bH\mathbf P+\kappa_0(\bA+\bm\Lambda_0) .
\end{split}
\label{eq:springs}
\end{equation}
Divided by $\kappa_0$, this is the DLS fit with $\kappa_0^{-1}\mathbf P\bH\mathbf P$ added to its regularizer, hence the
name: each deformation is penalized by its stiffness, and the structure gives way along its soft modes.

The restraint strength $\kappa_0$ sets the balance between the two terms (Fig.~\ref{fig:hessfit}a). For each source of structures, $\kappa_0$ is chosen the same way, by cross-validation on its pairs that are in no test set: the lowest shape error for the cell and, for the fit of the atoms (below), the lowest bond-length error at the
experimental cell (Supplementary Tables~11 and~12). This gives $\kapK$~eV for the cell and $\kapBK$~eV for the atoms of the DFT correction, and $\mlipKtece$~eV for the cells
and $\mlipKBtece$~eV for the atoms of the TECE-OAM-RRA-1.0 and EquFlashV2-45M-OAM corrections. Minimizing the full energy of MACE-OMAT-0 instead of its quadratic
expansion does not lower the shape error in cross-validation (Supplementary
Section~1).

\emph{Hessian-regularized DLS fit with a strain prior (estimator~5).} Estimator~5 adds to this fit a direct head of
its own, whose strain $\hat\bE_{\rm d}$
(equation~\eqref{eq:A}) acts as a prior on the cell: with a restraint strength $\kappa$ set for each structure in place of
$\kappa_0$ and the term $\tfrac12(\bq-\bq_{\rm d})^{\mathsf T}\bm\Lambda_{\rm d}(\bq-\bq_{\rm d})$ added to the objective
of equation~\eqref{eq:springs}, the minimum solves
\begin{equation}
\big(\mathbf P\bH\mathbf P+\kappa(\bA+\bm\Lambda_0)+\bm\Lambda_{\rm d}\big)\hat\bq=\kappa\mathbf b+\bm\Lambda_{\rm
d}\bq_{\rm d},\qquad \bm\Lambda_{\rm d}=\mathrm{diag}\big(\lambda_V\mathbf P_V+\lambda_S\mathbf P_S,\,\mathbf 0\big),
\label{eq:U}
\end{equation}
with $\bq_{\rm d}=(\gam(\hat\bE_{\rm d}),\mathbf 0)$ and $\mathbf P_V$ and $\mathbf P_S$ the matrices that
measure the volume and the shape of a strain to first order ($\gam^{\mathsf T}\mathbf P_V\gam=(\tr\bE)^2/3$, $\gam^{\mathsf
T}\mathbf P_S\gam=\|\dev\bE\|_F^2$, $\mathbf P_V+\mathbf P_S=\mathbf G_F$). The weights $\lambda_{V,S}$ and the strength $\kappa$ are scales times exponentials of bounded outputs of an MLP,
and the other estimators are limits of equation~\eqref{eq:U} (Supplementary Section~7).

\emph{Mean strain.} Each estimator is trained from three random initializations, and with $\hat\bE_r$ the mean of estimator $r$ over them,
the predicted strain is
\begin{equation}
\hat\bE=\frac15\sum_{r=1}^5\hat\bE_r .
\label{eq:mix}
\end{equation}
An estimator output or mean whose $\I+2\bE$ has an eigenvalue at or below $10^{-6}$ is evaluated as the uncorrected
structure; the means are taken over the raw outputs.
All five estimators use the MACE-OMAT-0 descriptors and the relaxation strains, and estimators~4 and 5 also the Hessian; Supplementary Table~5 compares this combination with simpler ones on the test sets.

\subsection*{Atomic positions}
\label{sec:m_atoms}
The atoms are placed by a separate Hessian-regularized DLS fit (Fig.~\ref{fig:arch}c), whose networks use the
MACE-OMAT-0 descriptors and the relaxation strains. Minimizing the objective of equation~\eqref{eq:springs} over $\bu$ at a
fixed cell $\gam$ gives
\begin{equation}
\hat\bu(\gam)=\mathbf M_{uu}^{-1}\big(\kappa_0\mathbf b_u-\mathbf M_{u\gamma}\gam\big),
\label{eq:clamped}
\end{equation}
with $\mathbf M_{uu}$ and $\mathbf M_{u\gamma}$ the displacement--displacement and displacement--strain blocks of
$\mathbf M$ and $\mathbf b_u$ the displacement part of $\mathbf b$:
each atom moves until the forces of the missing energy on it balance the restoring forces of the Hessian. The fit places
all atoms, including hydrogen. The displacements are evaluated at the predicted cell $\hat\gam=\gam(\hat\bE)$, averaged over three
initializations and combined with $\hat\bE$ by equation~\eqref{eq:pred}; for the comparisons in Results they are also
computed at the experimental cell. A new structure is corrected with the networks of all ten splits, $180$ in all,
whose strains and displacements are averaged (Supplementary Section~8).

\emph{Exact symmetry.}\label{sec:m_sym} All inputs are invariant and all
constructions above equivariant under rotations and reflections, so every space-group operation of the calculated
structure maps the prediction onto itself; all pairs within $r_c$ are used, since a list cut at a number of neighbours
could break the symmetry (Supplementary Section~3).

\subsection*{Training, model selection and evaluation}
\label{sec:m_training}
For each split, each estimator is trained from three random initializations. The losses act on the strain of the cell and on the bond lengths and displacements of the atoms, with no weight on hydrogen atoms (settings in Supplementary Tables~13 and~14). The predicted changes of the pair lengths have no target of their own; they are learned through the fits, whose linear solves pass the gradients of the losses back to the networks.

The inputs, estimators and
hyperparameters were chosen in cross-validation on the $\Ndev$ pairs that are in no test set (Supplementary Section~7).
The reported errors come from ten random splits ($\Ntrainmin$ training, $\Nvalmin$ validation and
$\Ntestmin$ test pairs) and two families of five folds grouped by structure prototype (anonymized formula,
space group and Wyckoff letters of the calculated structure)~\cite{mehl2017_aflow} or chemical system (set of elements),
each group held out in one fold (Supplementary Section~1). They leave out the $\Nomat$ pairs whose experimental structure is in OMat24 (except in the symmetry checks,
Supplementary Section~3). A reported value is the mean
of the ten test-set means, and a comparison gives the number of splits in which a model is better and a $95\%$
bootstrap interval of the relative change over structure prototypes~\cite{field2007_clusterboot} (Supplementary
Section~1).

\subsection*{Overlap with pretraining data and relaxation history}
\label{sec:m_exposure}
MPtrj, the relaxation trajectories of MP~\cite{deng2023_chgnet}, holds a frame that matches the experimental structure for $52\%$ of the pairs, so
potentials trained on it, such as TECE-OAM-RRA-1.0~\cite{tece2026} and
EquFlashV2-45M-OAM, supply no input; the input
potentials are trained on OMat24 alone. Because most MP relaxations
stop at an energy criterion, a calculated structure may retain part of its starting
geometry, which the correction
could exploit. To test this, calculated structures of pairs in no test set were relaxed again with the
Vienna Ab initio Simulation
Package (VASP)~\cite{kresse1996_prb,kresse1996_cms} to $0.005$~eV\,\AA$^{-1}$. In a
random sample, $\memIn$ converged
minima have a larger DFT shape error than the MP inputs ($\memRSdfttight$ against
$\memRSdftinp\%$), and models trained
without these pairs remove a similar share of it ($\memRSredtight$ against
$\memRSredinp\%$), although the corrected
errors there are higher, by $\memIVrel$, $\memISrel$ and $\memIBrel\%$ in volume, shape
and bond length
(Supplementary Section~9). The correction is thus calibrated to the
MP protocol, the setting of the
materials databases it is meant for.

\subsection*{Software}
\label{sec:m_software}
The code uses PyTorch~\cite{paszke2019_pytorch}, mace-torch, torch-dftd, fairchem-core, the Atomic Simulation Environment (ASE)~\cite{larsen2017_ase},
pymatgen~\cite{ong2013_pymatgen} and spglib~\cite{togo2018_spglib} (versions in Supplementary Section~7).

\backmatter

\bmhead{Supplementary information}
The Supplementary Information contains further results on the test splits, with the ablations and the size of the shape correction, the model selection, the symmetry checks, the targets and error measures, the data and their curation, the inputs, the networks, losses and training settings, the prediction for new structures, and the overlap with pretraining data and the relaxation history of the DFT structures, with the statistics and the software versions (Supplementary Sections~1--9
and Supplementary Tables~1--14).

\bmhead{Data availability}
The Materials Project and COD identifiers of the $\Nbench$ structure pairs, with their assignment to the splits and folds, are provided as Supplementary Data 1.
For each listed material, the calculated structure is its PBE or PBE+$U$ structure in MP (database version 2026.09.28), of the run type that MP's mixing rule assigns to its composition, and the experimental structure is the listed COD entry.

\bmhead{Code availability}
The code for training and evaluating RECAST will be available on GitHub at \url{https://github.com/utksi/recast} upon acceptance. The trained networks are available on Hugging Face at \url{https://huggingface.co/utksi/recast-models}. \linebreak
A web application that takes a calculated crystal structure, for example a VASP POSCAR or a CIF file, and returns the predicted experimental structure is available at: \url{https://huggingface.co/spaces/utksi/recast}.

\bmhead{Acknowledgements}
Computational resources were provided by the National Academic Infrastructure for Supercomputing in Sweden (NAISS and
LUMI), funded by the Swedish Research Council.

\bmhead{Funding}
U.S. acknowledges funding support from the Swedish e-Science Research Centre (SeRC). P.C.C. acknowledges support from the Knut and Alice Wallenberg Foundation (KAW-Scholar program). O.E. and E.K.D.-Cz. acknowledge
support from the Wallenberg Initiative Materials Science for Sustainability (WISE) funded by the Knut and Alice
Wallenberg Foundation (KAW), STandUPP, and eSSENCE. O.E. also acknowledges financial support from the Swedish Research
Council (VR), the European Research Council through the ERC Synergy Grant 854843-FASTCORR and the Knut and Alice
Wallenberg Foundation (KAW-Scholar program). S.I.S. acknowledges support from the Swedish Research Council (VR)
(2023-05247).
U.S, S.I.S., E. K. D.-Cz. and O.E acknowledge the support from Olle Engkvists Stiftelse (Grant No. 248-0422).

\FloatBarrier
\bmhead{Author contributions}

U.S. developed the structure-correction formalism, collected and curated the data, implemented and trained the models, and wrote the first draft of the manuscript.

P.C.C. and E.K.D.-Cz. collected and curated the data, and co-authored the manuscript. 

O.E. and S.I.S. conceived the project, and co-authored the manuscript.

S.I.S. proposed expressing the correction as a strain. 

All authors discussed the approach and results and contributed to the manuscript.

\bmhead{Competing interests}
The authors declare no competing interests.

\bibliography{references}

\end{document}


\nocite{hohenberg1964,kohn1965,pbe1996,kurth1999_errorpbe,Lejaeghere2014_errorpbe,isaacs2018,lejaeghere2016,jain2013,dudarev1998,chen2022_m3gnet,deng2023_chgnet,batatia2022_mace,batatia2024_macemp0,kellner2026,mannan2026_uniffbench,sun2015_scan,furness2020_r2scan,kingsbury2022_r2scan,balabin2009,ramakrishnan2015,hu2003,simak2025,gong2022,hussein2022,yuk2024,vandewalle1999,rocken2024,yang2024_deeprelax,yang2026_e3relax,fogarasi1992,melli2021,lemm2021_g2s,yoon2020_dogss,senior2020_alphafold,yang2020_trrosetta,hiranuma2021,jack1978,zubatyuk2025,wallace1972,grazulis2012_cod,meier1969,barrosoluque2024_omat24,grimme2010_d3,tece2026,equflashv2_ckpt,riebesell2025_matbench,matbench_leaderboard,allen1986,woinska2016,zagorac2019_icsd,fu2025_esen,grimme2011_bj,larsen2019_dim,wu2005_relaxedion,dejong2015_elastic,klipfel2023,zeni2025_mattergen,waser1963_restraints,mehl2017_aflow,field2007_clusterboot,kresse1996_prb,kresse1996_cms,paszke2019_pytorch,larsen2017_ase,ong2013_pymatgen,togo2018_spglib}

\begin{center}
{\Large\bfseries Supplementary Information:}\\[4pt]
{\Large \bfseries RECAST: Approaching experimental accuracy in theoretical calculations of crystal structures}\\[4pt]
Utkarsh Singh, Pamela Costa Carvalho, Erna K.\ Delczeg-Czirjak, Olle Eriksson, Sergei I.\ Simak
\end{center}

\noindent Equations numbered with an S are in this document; all other equations, the figures, the symbols and the
reference numbers are those of the main text (references cited only here continue its list). Abbreviations:
density-functional theory (DFT), the Perdew--Burke--Ernzerhof functional (PBE), DFT and PBE with an on-site Hubbard
correction (DFT+$U$, PBE+$U$), Materials Project (MP), Crystallography Open
Database (COD), machine-learned interatomic potential (MLIP), equivariant smooth energy network (eSEN),
distance-least-squares (DLS), Open Materials 2024 (OMat24), the relaxation trajectories of MP (MPtrj), the D3 dispersion
correction (D3), the fast inertial relaxation engine (FIRE), the Atomic Simulation Environment (ASE) and the
r$^2$SCAN meta-generalized-gradient approximation.

\tableofcontents

\section{Evaluation and further results}
\label{si:results}
\subsection*{Evaluation and statistics}

All models are trained and evaluated on ten random splits of the structure pairs, each with $\Ntrainmin$ training,
$\Nvalmin$ validation and $\Ntestmin$ test pairs, and on two
families of five folds that hold out whole structure prototypes or chemical systems (\emph{Grouped folds} below). A test
prediction is one pair in the test set of one split. The $\Nomat$ pairs whose experimental structure appears in OMat24
(Supplementary Section~\ref{si:overlap}) are left out of all reported errors; they enter only the symmetry checks of
Supplementary Section~\ref{si:sym}.

Reported values are means of the ten test-set means. Because the test sets overlap, the spread between splits is not a
confidence interval. For comparisons, each distinct test pair is first averaged over the splits that test it, and the
relative change of the mean error over the $\Npairsrand$ distinct pairs is bootstrapped over whole structure prototypes
($10{,}000$ resamples)~\cite{field2007_clusterboot}; the two-sided $95\%$ percentile intervals reflect the sampling of
structures, not the variation between training runs. Values by stratum are pooled over the test predictions of the ten
splits.

\subsection*{Accuracy}

Supplementary Table~\ref{tab:si_errors} lists all error measures, and Supplementary Table~\ref{tab:si_split} gives
those of the cell and of the atoms other than hydrogen for each split. The correction is more accurate than DFT in each split and for each measure, except for the bonds
to hydrogen, which improve in $\XHk$ of the ten splits (there are only $\XHn$ neutron structures among the test
pairs).

\begin{table}[ht]
\centering
\small
\setlength{\tabcolsep}{3.5pt}
\caption{Errors of calculated and corrected structures against experiment. Means over the ten test sets ($e_{XH}$:
over the test predictions of the $\XHn$ neutron structures). Uncorrected: the calculated structure. Constant: one isotropic scaling per
split, the median volume ratio of its training structures, with the atoms at their fractional coordinates. Learned: the
correction trained for that source, with hyperparameters chosen in model selection (Supplementary Section~\ref{si:train}). Volume $e_V$, cell shape $e_S$, lattice-vector lengths $e_L$, cell angles $e_A$, bond
lengths $e_B$, atomic displacements $e_D$ (root mean square, rigid translation removed) and bond angles $e_\theta$ of the atoms other than hydrogen, and bonds
to hydrogen $e_{XH}$ (equation~(4) of the main text).}
\label{tab:si_errors}
\begin{tabular}{llcccccccc}
\toprule
Source & Correction & $e_V$ (\%) & $e_S$ (\%) & $e_L$ (\%) & $e_A$ ($^\circ$) & $e_B$ (\%) & $e_D$ (\AA) & $e_\theta$ ($^\circ$) & $e_{XH}$ (\%)\\
\midrule
DFT & uncorrected & $\eVPBE$ & $\eSPBE$ & $\eLPBE$ & $\eAPBE$ & $\eBPBE$ & $\eDPBE$ & $\eTPBE$ & $\XHPBE$\\
  & constant & $\constVdft$ & $\constSdft$ & $\constLdft$ & $\constAdft$ & $\constBdft$ & $\constDdft$ & $\constTdft$ & $\constXHdft$\\
 & learned & $\eVFin$ & $\eSFin$ & $\eLFin$ & $\eAFin$ & $\eBFin$ & $\eDFin$ & $\eTFin$ & $\XHFin$\\
\midrule
TECE-OAM-RRA-1.0 & uncorrected & $\teceV$ & $\teceS$ & $\teceL$ & $\teceA$ & $\teceB$ & $\teceD$ & $\teceT$ & $\teceXH$\\
 & constant & $\constVtece$ & $\constStece$ & $\constLtece$ & $\constAtece$ & $\constBtece$ & $\constDtece$ & $\constTtece$ & $\constXHtece$\\
 & learned & $\teceVFin$ & $\teceSFin$ & $\teceLFin$ & $\teceAFin$ & $\teceBFin$ & $\teceDFin$ & $\teceTFin$ & $\teceXHFin$\\
\midrule
EquFlashV2-45M-OAM & uncorrected & $\eqfV$ & $\eqfS$ & $\eqfL$ & $\eqfA$ & $\eqfB$ & $\eqfD$ & $\eqfT$ & $\eqfXH$\\
 & constant & $\constVeqf$ & $\constSeqf$ & $\constLeqf$ & $\constAeqf$ & $\constBeqf$ & $\constDeqf$ & $\constTeqf$ & $\constXHeqf$\\
 & learned & $\eqfVFin$ & $\eqfSFin$ & $\eqfLFin$ & $\eqfAFin$ & $\eqfBFin$ & $\eqfDFin$ & $\eqfTFin$ & $\eqfXHFin$\\
\bottomrule
\end{tabular}
\end{table}

\begin{table}[ht]
\centering
\small
\caption{Errors per split $p$. Top, the cell: number of test predictions evaluated, volume error $e_V$ (\%), shape
error $e_S$ (\%) of DFT, of the correction and of the five estimators without the relaxation strains $\bT_k$ (``no
$\bT_k$''; one initialization each), lattice-vector-length error $e_L$ (\%) and cell-angle error $e_A$ ($^\circ$).
Bottom, the atoms: bond lengths $e_B$ (\%) of DFT, of the experimental cell with the DFT fractional coordinates
($^a$) and with the predicted displacements ($^b$), and of the corrected structure; displacement error $e_D$ (\AA)
and bond-angle error $e_\theta$ ($^\circ$) of DFT and of the corrected structure.}
\label{tab:si_split}
\begin{tabular}{lrcccccccccc}
\toprule
 & & \multicolumn{2}{c}{$e_V$} & \multicolumn{3}{c}{$e_S$} & \multicolumn{2}{c}{$e_L$} & \multicolumn{2}{c}{$e_A$}\\
\cmidrule(lr){3-4}\cmidrule(lr){5-7}\cmidrule(lr){8-9}\cmidrule(lr){10-11}
$p$ & $n$ & DFT & corrected & DFT & corrected & no $\bT_k$ & DFT & corrected & DFT & corrected\\
\midrule
\tabpartcell
\bottomrule
\end{tabular}

\vspace{10pt}

\begin{tabular}{lcccccccc}
\toprule
 & \multicolumn{4}{c}{$e_B$} & \multicolumn{2}{c}{$e_D$} & \multicolumn{2}{c}{$e_\theta$}\\
\cmidrule(lr){2-5}\cmidrule(lr){6-7}\cmidrule(lr){8-9}
$p$ & DFT & exp.\ cell$^a$ & exp.\ cell$^b$ & corrected & DFT & corrected & DFT & corrected\\
\midrule
\tabpartatoms
\bottomrule
\end{tabular}
\end{table}

Resampling whole chemical systems instead of structure prototypes gives nearly the same
intervals of the relative changes from DFT as in Fig.~2 of the main text:
$\randCVrello$ to $\randCVrelhi\%$ (volume), $\randCSrello$ to $\randCSrelhi\%$ (shape) and $\randCBrello$ to
$\randCBrelhi\%$ (bonds). Leaving out near duplicates, test pairs whose reduced formula and structure prototype both occur
among the training or validation pairs of their split ($\nodupremoved$ per split on average), changes the mean errors of
DFT and of the correction by at most $0.01$ percentage points.

\subsection*{Simpler corrections}

Supplementary Table~\ref{tab:si_simple} compares the correction with four simpler ones, each on the test
predictions where it is defined: the constant correction of Supplementary Table~\ref{tab:si_errors}, a hydrostatic
pressure and the cells relaxed with MACE-OMAT-0 with and without D3 (Supplementary Section~\ref{si:inputs}). The
pressure brings each cell to its experimental volume through the relaxed-ion stiffness $\mathbf C$ of MACE-OMAT-0,
with the strain $\bE_p$ of Voigt vector $\gam_p$,
\begin{equation}
\gam_p\propto\mathbf C^{-1}(1,1,1,0,0,0)^{\mathsf T},\qquad \det(\I+2\bE_p)=\det(\I+2\bE^*),
\label{eq:si_hyd}
\end{equation}
for the pairs whose $\mathbf C$ is positive definite. It removes the volume error by construction but raises the
shape error, so a missing pressure does not explain the anisotropy of the DFT error. The relaxed cells have larger
shape errors than DFT, those relaxed with D3 at a smaller volume error and those without at a larger one; their
strains enter the correction only as inputs.

\begin{table}[ht]
\centering
\small
\caption{Simpler corrections. Test predictions $n$ on which each is defined, and the volume error $e_V$ and shape
error $e_S$ (\%) of DFT, of the simpler correction and of the correction on the same predictions; means over the ten
test sets. Constant correction: one isotropic scaling per split. Hydrostatic pressure: equation~\eqref{eq:si_hyd},
for the pairs whose relaxed-ion stiffness in MACE-OMAT-0 is positive definite. Relaxations: the cells relaxed with
MACE-OMAT-0 with and without D3, where the relaxation converged and kept the space group of the calculated
structure.}
\label{tab:si_simple}
\begin{tabular}{lrcccccc}
\toprule
 & & \multicolumn{3}{c}{$e_V$ (\%)} & \multicolumn{3}{c}{$e_S$ (\%)}\\
\cmidrule(lr){3-5}\cmidrule(lr){6-8}
 & $n$ & DFT & simpler & corrected & DFT & simpler & corrected\\
\midrule
\tabsimple
\bottomrule
\end{tabular}
\end{table}

\subsection*{Pretraining data of the MLIPs}

MPtrj, on which TECE-OAM-RRA-1.0 and EquFlashV2-45M-OAM were pretrained,
contains the experimental structure of $\expteceInn$ of the distinct test pairs and not that of $\expteceOutn$; Supplementary
Table~\ref{tab:si_exposure} gives the errors of both groups.

\begin{table}[ht]
\centering
\small
\caption{Errors by overlap with the pretraining data of the MLIPs. Mean errors (\%) of the distinct test pairs, each averaged over the splits that test it, by whether an MPtrj
frame is the pair's experimental structure (exact structure match); the $\expnorec$ test pairs whose material has no
MPtrj record are in neither group.}
\label{tab:si_exposure}
\begin{tabular}{llcccccc}
\toprule
 & & \multicolumn{2}{c}{$e_V$} & \multicolumn{2}{c}{$e_S$} & \multicolumn{2}{c}{$e_B$}\\
Source & In MPtrj & uncorr. & corr. & uncorr. & corr. & uncorr. & corr.\\
\midrule
DFT & yes ($\expdftInn$) & $\expdftInVraw$ & $\expdftInVfin$ & $\expdftInSraw$ & $\expdftInSfin$ & $\expdftInBraw$ & $\expdftInBfin$\\
 & no ($\expdftOutn$) & $\expdftOutVraw$ & $\expdftOutVfin$ & $\expdftOutSraw$ & $\expdftOutSfin$ & $\expdftOutBraw$ & $\expdftOutBfin$\\
TECE-OAM-RRA-1.0 & yes & $\expteceInVraw$ & $\expteceInVfin$ & $\expteceInSraw$ & $\expteceInSfin$ & $\expteceInBraw$ & $\expteceInBfin$\\
 & no & $\expteceOutVraw$ & $\expteceOutVfin$ & $\expteceOutSraw$ & $\expteceOutSfin$ & $\expteceOutBraw$ & $\expteceOutBfin$\\
EquFlashV2-45M-OAM & yes & $\expeqfInVraw$ & $\expeqfInVfin$ & $\expeqfInSraw$ & $\expeqfInSfin$ & $\expeqfInBraw$ & $\expeqfInBfin$\\
 & no & $\expeqfOutVraw$ & $\expeqfOutVfin$ & $\expeqfOutSraw$ & $\expeqfOutSfin$ & $\expeqfOutBraw$ & $\expeqfOutBfin$\\
\bottomrule
\end{tabular}
\end{table}

\subsection*{Ablations and learning curve}

To see what each part of the correction contributes on unseen data, ablation
networks were trained on the same splits with the settings of the model: two more initializations of estimators~1
and 5 and, on five splits, estimators 1 to 5 and four more initializations of estimator~5 on nested random subsets of
$25$ and $50\%$ of the training pairs, with unchanged validation and test pairs. Each variant is the mean strain of the
listed networks, with the atoms placed by the atom fit of the model (Supplementary Table~\ref{tab:si_testabl}).

\begin{table}[ht]
\centering
\footnotesize
\setlength{\tabcolsep}{4pt}
\caption{Ablations on the test sets. Mean errors over the test sets of the ten splits (top) and of five splits with
networks trained on nested random subsets of the training pairs (bottom; validation and test pairs unchanged): volume
$e_V$, shape $e_S$, lattice-vector lengths $e_L$ (\%), cell angles $e_A$ ($^\circ$) and bond lengths $e_B$ (\%), with
the atoms of the model at each cell. Estimators as in the main text: 1, the direct strain head; 2 and 3, the DLS
fits of the cell and of the cell and atoms; 4, the Hessian-regularized DLS fit; 5, the same with a strain prior.
Networks: the number of trained networks averaged.}
\label{tab:si_testabl}
\begin{tabular}{lcccccc}
\toprule
Cell estimate & Networks & $e_V$ & $e_S$ & $e_L$ & $e_A$ & $e_B$\\
\midrule
\tabrefAsi
\midrule
\tabrefB
\bottomrule
\end{tabular}
\end{table}

\subsection*{Strata, the largest errors and the size of the shape correction}

Supplementary Table~\ref{tab:si_strata} gives the errors by stratum: the DFT shape error, the dimensionality of the
bonded network, a change of space group, high-pressure and high- or low-temperature phases, the functional and the
size of the cell.

\begin{table}[htbp]
\centering
\small
\setlength{\tabcolsep}{3pt}
\caption{Errors by stratum. Means pooled over the $n$ test predictions of the ten splits: shape error $e_S$ and volume error
$e_V$ (\%) of DFT, of the five estimators with one initialization each, without and with the relaxation strains
(``no $\bT_k$'' and ``$\bT_k$''), and of the correction (three initializations each), and
bond-length error $e_B$ (\%) of DFT and of the corrected structure. $s_{\rm DFT}$ is the DFT shape error. Space group: spglib groups of the DFT and the experimental
structure at a tolerance of $10^{-3}$~\AA. High-pressure and high- or low-temperature ($T$) phases: so marked in the tags or
remarks of the MP records. Functional: DFT and DFT+$U$. Atoms: the number of atoms in
the cell of the DFT structure.}
\label{tab:si_strata}
\begin{tabular}{lrcccccccccc}
\toprule
 & & \multicolumn{4}{c}{$e_S$} & \multicolumn{4}{c}{$e_V$} & \multicolumn{2}{c}{$e_B$}\\
\cmidrule(lr){3-6}\cmidrule(lr){7-10}\cmidrule(lr){11-12}
Stratum & $n$ & DFT & no $\bT_k$ & $\bT_k$ & corrected & DFT & no $\bT_k$ & $\bT_k$ & corrected & DFT & corrected\\
\midrule
\tabstratablocks
\bottomrule
\end{tabular}
\end{table}

The pairs with a DFT shape error above $3\%$ contain hydrogen more often ($\tailH$ against $\allH\%$ of all pairs)
and are softer: their median relaxed-ion shear modulus in MACE-OMAT-0 is $\tailshear$~GPa, against $\allshear$~GPa for
all pairs.

\label{si:shape}Where the DFT error is large, the predicted shape changes point the right way but are too small. In the
out-of-fold predictions on the model-selection set ($\shpN$ structures, one initialization per estimator, as in
Supplementary Table~\ref{tab:si_potential}), the $\shpTailN$ structures with a DFT shape error of at least $3\%$ carry
$\shpTailShare\%$ of the corrected shape error; for them the predicted traceless strain points along the needed one
(median cosine $\shpCos$) but has a median of $\shpRatio$ of its size. The quadratic model of the fit does not explain
the missing size. Minimizing the objective of the Hessian-regularized DLS fit with the full energy of MACE-OMAT-0 in
place of its quadratic model does not lower the shape error of that estimator or of the five-estimator mean, and for
the structures above $3\%$ it raises the error of that estimator from $\shpNLsprTailLin$ to $\shpNLsprTailNl\%$: the
energy surface stiffens under compression and shrinks the predicted change of shape, by a median of $\shpNLalong\%$.
The minimization was run for $\shpNLn$ structures of the model-selection set (all those with a DFT shape error of at least $3\%$ and
a random sample of the others), with FIRE and the cell filter, in the space group of each structure, to forces below
$2.5\times10^{-4}$~eV\,\AA$^{-1}$.

\subsection*{Atomic positions}

Supplementary Table~\ref{tab:si_ladder} separates the errors of the atomic structure into the parts due to the cell
and to the positions within it.

\begin{table}[ht]
\centering
\small
\caption{Errors of the atomic structure when part of the experimental structure is supplied. Means over the ten test
sets; $e_{XH}$ for the $\XHn$ neutron structures. Each row combines a cell with a choice of atomic positions: DFT
fractional coordinates keep each atom at its DFT position relative to the lattice vectors; predicted displacements are
those of the Hessian-regularized DLS fit at the cell of the row, including hydrogen. Rows two to four use the experimental volume or cell and are
diagnostics, not predictions; the last row is the corrected structure.}
\label{tab:si_ladder}
\begin{tabular}{llcccc}
\toprule
Cell & Atomic positions & $e_B$ (\%) & $e_D$ (\AA) & $e_\theta$ ($^\circ$) & $e_{XH}$ (\%)\\
\midrule
DFT & DFT & $\eBPBE$ & $\eDPBE$ & $\eTPBE$ & $\XHPBE$\\
DFT, scaled to $V^*$ & DFT fractional coordinates & $\eBIso$ & $\eDIso$ & $\eTIso$ & $\XHIso$\\
experimental & DFT fractional coordinates & $\eBMeas$ & $\eDPBE$ & $\eTMeas$ & $\XHMeas$\\
experimental & predicted displacements & $\eBMeasHyb$ & $\eDMeasHyb$ & $\eTMeasHyb$ & $\XHMeasHyb$\\
predicted & predicted displacements & $\eBFin$ & $\eDFin$ & $\eTFin$ & $\XHFin$\\
\bottomrule
\end{tabular}
\end{table}

To judge the predicted displacements apart from the cell, the atoms of the
$\TgcN$ test pairs with displacement targets (without the pairs whose experimental structure is in OMat24) are placed in the
experimental cell. The bond-length error $e_B$, a mean over these pairs, each averaged over the splits that test it, is then $\TgcBcell\%$ with the DFT fractional coordinates,
$\TgcBmodel\%$ with the predicted displacements and $\TgcBrelax\%$ after relaxing the atoms in that cell with MACE-OMAT-0
(positions only, symmetry kept, forces below $0.01$~eV\,\AA$^{-1}$), which moves them towards the minimum of the
potential rather than towards experiment. At the predicted cell the three values are $\TgcPBcell$, $\TgcPBmodel$ and
$\TgcPBrelax\%$. Supplementary Table~\ref{tab:si_ladder} gives the means of the ten test-set means instead.
A bond-tension network (Supplementary Section~\ref{si:train}; one initialization) that moves the atoms
through the Hessian of MACE-OMAT-0 gives a bond-length error of $\rAtB\%$ at the experimental cell; the same network
predicting the displacements directly gives $\rAtBdir\%$ and is worse in all ten splits.

\subsection*{Grouped folds}
\label{si:grouped}

To test transfer to structure types and chemistries absent from training,
the pairs are also grouped by structure prototype, given by the anonymized formula (the sorted reduced stoichiometry, such
as AB$_2$), the space-group number and the sorted Wyckoff letters of the calculated structure (spglib, tolerance
$10^{-3}$~\AA)~\cite{mehl2017_aflow}, or by chemical system, the set of elements. The $\Nbench$ pairs form $7{,}432$
prototypes, $5{,}145$ of them with a single pair and the largest, the ThCr$_2$Si$_2$ type, with $252$, and $11{,}626$
chemical systems, the largest (O--Si) with $44$ pairs. In each family the groups are assigned, largest first, to the fold
with the fewest pairs, which gives five folds of $3{,}611$ or $3{,}612$ pairs. Each fold is predicted by a model trained on
the other four, with $10\%$ of their pairs for early stopping, so that it trains on $72\%$ of all pairs, against $81\%$ in
the random splits; the model and its settings are those of the random splits. A held-out group takes all its pairs with
it but not those of related groups: with O--Si held out, Li--O--Si pairs can be in training, and so can pairs that
differ from a held-out prototype in formula type, space group or Wyckoff sites.

\begin{table}[ht]
\centering
\small
\setlength{\tabcolsep}{3.5pt}
\caption{Errors on the grouped folds. In both families, each structure prototype (or chemical system) is held out once,
in one of five folds, so that each pair is predicted once ($\gPpairs$ pairs) by a model that saw no pair of its
group; means over the pairs. All seven errors are lower than for DFT in each of the five folds of both
families. Constant: one isotropic scaling per fold, the median volume ratio of its training pairs.}
\label{tab:si_grouped}
\begin{tabular}{llccccccc}
\toprule
Held out & Correction & $e_V$ (\%) & $e_S$ (\%) & $e_L$ (\%) & $e_A$ ($^\circ$) & $e_B$ (\%) & $e_D$ (\AA) & $e_\theta$ ($^\circ$)\\
\midrule
prototype & DFT & $\gPVref$ & $\gPSref$ & $\gPLref$ & $\gPAref$ & $\gPBref$ & $\gPDref$ & $\gPTref$\\
 & constant & $\gPVconst$ & $\gPSconst$ & $\gPLconst$ & $\gPAconst$ & $\gPBconst$ & $\gPDconst$ & $\gPTconst$\\
 & learned & $\gPVcorr$ & $\gPScorr$ & $\gPLcorr$ & $\gPAcorr$ & $\gPBcorr$ & $\gPDcorr$ & $\gPTcorr$\\
\midrule
chemical system & DFT & $\gCVref$ & $\gCSref$ & $\gCLref$ & $\gCAref$ & $\gCBref$ & $\gCDref$ & $\gCTref$\\
 & constant & $\gCVconst$ & $\gCSconst$ & $\gCLconst$ & $\gCAconst$ & $\gCBconst$ & $\gCDconst$ & $\gCTconst$\\
 & learned & $\gCVcorr$ & $\gCScorr$ & $\gCLcorr$ & $\gCAcorr$ & $\gCBcorr$ & $\gCDcorr$ & $\gCTcorr$\\
\bottomrule
\end{tabular}
\end{table}

\section{Model selection}
\label{si:dev}

All design choices were made on the model-selection set, the $\Ndev$ pairs that are in no test set
($\Ndevscored$ of them are evaluated; the others have their experimental structure in OMat24). For each choice, a few options were compared on ten folds
of this set, and a change was adopted only if it lowered the mean shape error across the folds without
raising the volume error, according to the paired test of equation~\eqref{eq:ttest} with Holm's correction~\cite{holm1979}
for the number of options. Hyperparameters and restraint strengths were set by searches on further folds (Supplementary
Section~\ref{si:train}). The test sets, the grouped folds and the ablations were evaluated once, after the model was
fixed, and none of them changed it.

MACE-OMAT-0 and eSEN-30M-OMat, both trained on OMat24 alone, were compared as sources of the descriptors and of the
physics (the Hessian and the relaxation strains), in all four combinations on the model-selection set, with the cell
estimators and the atom fit retrained for each (Supplementary Table~\ref{tab:si_potential}). Taking everything from
eSEN-30M-OMat would lower the test errors by $\potTestV\%$ in volume and $\potTestS\%$ in shape, but this potential
is equivariant under rotations and not under reflections: with its Hessian, $\potSymLose$ of the $\potSymN$ corrected
test structures with an unambiguous space group lose it at a tolerance of $10^{-3}$~\AA, against none with that of
MACE-OMAT-0, which therefore supplies the physics. Its descriptors are used as well: with this physics, those of
eSEN-30M-OMat do not lower the shape error in model selection.

\begin{table}[ht]
\centering
\small
\caption{Pretrained potentials. Model-selection set ($\potN$ structures, ten folds): the potential that supplies the
descriptors and the one that supplies the physics (Hessian and relaxation strains); mean over the folds of the fold
means, in percent; the cell estimators and the atom fit trained anew for each combination, $e_B$ of the atom fit at the
experimental cell. The correction takes its descriptors and its physics from MACE-OMAT-0.}
\label{tab:si_potential}
\begin{tabular}{@{}llccc@{}}
\toprule
Descriptors & Physics & $e_V$ & $e_S$ & $e_B$\\
\midrule
MACE-OMAT-0 & MACE-OMAT-0 & $\potMMV$ & $\potMMS$ & $\potMMB$\\
MACE-OMAT-0 & eSEN-30M-OMat & $\potMEV$ & $\potMES$ & $\potMEB$\\
eSEN-30M-OMat & MACE-OMAT-0 & $\potEMV$ & $\potEMS$ & $\potEMB$\\
eSEN-30M-OMat & eSEN-30M-OMat & $\potEEV$ & $\potEES$ & $\potEEB$\\
\midrule
DFT, uncorrected & & $\potDFTV$ & $\potDFTS$ & \\
\bottomrule
\end{tabular}
\end{table}

\section{Symmetry: exactness and numerical precision}
\label{si:sym}
\subsection*{Exact symmetry}

A rotation or reflection $\mathbf R$ of the calculated structure acts on the coordinates as $\bq\to\bq^{\mathbf R}$,
with $\gam^{\mathbf R}=\gam(\mathbf R\bE\mathbf R^{\mathsf T})$ and $\bu_k^{\mathbf R}=\bu_k\mathbf R^{\mathsf T}$, which
preserves $\|\bE\|_F$ and $\|\bu\|$. The descriptors and all network outputs are invariant under $\mathbf R$; the tensors
$\bT^{\rm b}_i$, $\bT_k$ and $\mathbf Q^\nu_i$ map as $\mathbf X\to\mathbf R\mathbf X\mathbf R^{\mathsf T}$; each
$\mathbf a_p\cdot\bq$ is unchanged when the structure and $\bq$ rotate together; and the Hessian, truncated in coordinates
in which the map is orthogonal, transforms with them. Each objective of the fits is therefore invariant, and its unique
minimizer equivariant (the matrices of equations~(10), (12) and (13) of the main
text are positive definite); the direct strain head is equivariant term by term. Hence
$\hat\bE\to\mathbf R\hat\bE\mathbf R^{\mathsf T}$ and $\hat\bu_k\to\hat\bu_k\mathbf R^{\mathsf T}$. Because the prediction
is also invariant under translations and equivariant under permutations of the atoms, every space-group operation of the
calculated structure maps it onto itself: in exact arithmetic the corrected structure keeps the space group of the
calculated one, provided that the set of pairs does. All pairs within $r_c$ are therefore used; a list cut at a fixed
number of neighbours inside a shell of equal distances would break the symmetry.

\subsection*{Numerical precision}

In double precision and with untrained networks, the fits reproduce the transformed predictions of rotated, reflected
and permuted structures and of supercells to $1.1\times10^{-12}$ relative or better, also with the Hessian
transformed with the structure. In the reported predictions the networks run in single precision; for the test
predictions with a DFT shape error below $10^{-3}\%$, nearly all of them cubic cells, the predicted shape change has a
median of $\cubmed\%$.

\subsection*{Symmetry of the test predictions}

The space group of the corrected structure was compared with that of the DFT structure, found by spglib at
$10^{-5}$~\AA, for each test pair ($\TsymN$, including those whose experimental structure is in OMat24). For the
$\TsymExact$
inputs whose space group is the same at $10^{-5}$ and $10^{-2}$~\AA, the corrected structures keep it at $10^{-3}$ and
$10^{-2}$~\AA\ in each case and at $10^{-5}$~\AA\ in all but $\TsymLosing$, which recover it after a projection that
moves the atoms by at most $\TsymProjMove$~\AA. The $\TsymApproxN$ inputs that reach a higher space group only at a
looser tolerance are corrected within the group found at $10^{-5}$~\AA; where the higher symmetry is intended, the
output can be projected onto that group.

\section{Targets, correspondence and error measures}
\label{si:labels}
\subsection*{Properties of the targets}
With the targets of equations~(2) and (3) of the main text,
$\bL(\bE^*)\mathbf R_0=\bL^*$ and $\br_k(\bE^*,\bu^*)\mathbf R_0=\mathbf f^*_k\bL^*-\mathbf t$ for all $k$ and one
translation $\mathbf t$.

\subsection*{Error measures}

The bonds $\mathcal B$ are the pairs $(i,j,\mathbf m)$ of the experimental structure between atoms other than hydrogen, with
$\mathbf m$ the lattice translation of atom $j$, whose length $d$ satisfies $d\le1.15\,d_i^{\min}$, where $d_i^{\min}$ is
the distance of atom $i$ to its nearest neighbour, including hydrogen; a bond in the first shell of both its atoms enters twice.
The errors are those of equation~(4) of the main text; $e_\theta$ is the mean absolute error of the
angles at each atom between two bonds of $\mathcal B$ that contain it and lead to different neighbours, each bond counted
as often as it enters $\mathcal B$, as in $e_B$. $e_B$ and $e_\theta$ are undefined for the $\NnoB$ evaluated structures in
which each atom other than hydrogen has only hydrogen in its first shell, and $e_\theta$ also for the $\NnoT$ structures in which no
atom has bonds to two different neighbours. $e_S$ does not depend on the orientation or the lattice basis; $e_L$ and
$e_A$ refer to the basis of the target. $e_{XH}$ is $e_B$ for the bonds to hydrogen, evaluated only for the neutron structures,
whose COD records give neutrons as the radiation of the diffraction experiment ($\Nneutron$ pairs, $\XHn$ of them in a test set). The dimensionality of the bonded network of a DFT structure is that of its
highest-dimensional bonded component, found with the method of Larsen et al.~\cite{larsen2019_dim} for bonds shorter
than $1.3$ times the sum of the covalent radii tabulated in ASE~\cite{larsen2017_ase}; it only groups the errors and is
not an input of the correction.

\subsection*{Targets and correspondence}

\paragraph{Lattice correspondence.} The two cells of a pair are compared in one lattice basis: the stored one, unless
another basis among the $6{,}960$ unimodular matrices $\mathbf N\in\{-1,0,1\}^{3\times3}$ ($|\det\mathbf N|=1$) gives a
strain smaller by more than $1\%$ in $\tfrac12\|\ln(\I+2\bE)\|_F$ and fits the atoms, over origins and assignments of
atoms of the same species, at most $0.02$~\AA\ (root mean square) worse. Of the $24$ such bases with the smallest strains,
the one that fits the atoms best is used; this applies to $\Nalt$ pairs.

\paragraph{Displacements and changed space groups.} The displacement target $\bu^*_k$ (equation~(3) of the
main text) uses the periodic image of each atom nearest its experimental position. It requires the stored basis, a
root-mean-square displacement of at most $0.3$~\AA\ and no displacement of an atom other than hydrogen above $1$~\AA.
 Two kinds of pair are identified: those whose atoms are displaced by more than $0.3$~\AA\ (root mean square) even under the
best match of origin, species and basis ($\Nrearr$ pairs) and those whose space group, found by spglib at $10^{-3}$~\AA, differs between the two structures
($\Nsg$; the curation of Supplementary Section~\ref{si:data} accepts the space group reported by the COD or found at up to
$0.1$~\AA). These pairs, $\idflag\%$ of all, stay in the benchmark;
leaving them out changes the mean test errors by about 0.01 percentage points at most.

\paragraph{Second experimental reports.} A second report of a pair is a COD entry of the same phase, that is, another entry matched with the same MP material
(Supplementary Section~\ref{si:data}), whose volume differs from that of the experimental structure of the pair by more
than $10^{-4}$ (relative), so that it is not a copy, and whose primitive cell holds as many atoms as that structure's; it need not come from a
separate publication, and it may have been
measured at another temperature. The shape difference between two reports is the smallest $\|\shp(\bE)\|_F$ between their
Niggli-reduced cells~\cite{krivy1976_niggli} over changes of basis, and their bond-length difference is $e_B$ after an
atom-by-atom match (pymatgen StructureMatcher~\cite{ong2013_pymatgen}; tolerances $0.2$ in the lattice lengths, $0.3$ in
the positions and $5^\circ$ in the angles) for the pairs with bond-length errors (those with displacement targets and bonds between atoms other than
hydrogen). With several second
reports, the comparison of the main text uses their median, over the distinct test pairs, each averaged over the splits
that test it, and the analysis of the largest DFT errors uses the closest report. The shape errors of other methods, such
as the r$^2$SCAN relaxations of MP, are compared with those of DFT as tensors in the frame of the experimental cell,
$\mathbf e_m=\tfrac12(\mathbf L_e^{\mathsf T}\bG_m^{-1}\mathbf L_e-\I)$ with $\bG^*=\mathbf L_e\mathbf L_e^{\mathsf T}$
and $\bG_m$ the metric tensor of method $m$, by the cosine of their traceless parts.

\paragraph{MLIP structures.} The MLIP structures of Fig.~2 of the main text are relaxations of the
experimental (COD) structure of each pair with the released models (TECE-OAM-RRA-1.0 version 2026.07.05,
EquFlashV2-45M-OAM version 2026.06.11). The space group of the input is kept with the FixSymmetry constraint of the
Atomic Simulation Environment (ASE)~\cite{larsen2017_ase} (tolerance $10^{-2}$~\AA); the cell and the positions are
relaxed with FIRE and a Fr\'echet cell filter, in the single precision of the released models, to forces of $0.01$~eV\,\AA$^{-1}$. A relaxation not converged after $500$ steps is
continued from its end point for at most $3{,}000$ further steps ($10\%$ of the structures); the $1\%$ still unconverged
are used as they are. Their targets are defined as for DFT, with the MLIP cell and positions in place of the DFT ones,
and the MLIP corrections are evaluated on the same test predictions.

\section{Structure pairs, data and curation}
\label{si:data}
\subsection*{Structure pairs}

The MP data are those of database version 2026.09.28, retrieved on 10 and 11 November 2026; the composition of the benchmark is
given in Supplementary Table~\ref{tab:si_pop}. For the MLIP corrections, the calculated structure of a pair is the
relaxation of its experimental structure by the MLIP (Supplementary Section~\ref{si:labels}). The r$^2$SCAN
structure of a material, compared with DFT in the main text, is the final structure of an r$^2$SCAN structure
optimization in MP. It is the one MP designates for the material (its blessed task) or, where that is not a completed
structure optimization, the completed one with the smallest residual pressure. Such a structure exists for
$\allrscann$ pairs.

\subsection*{Data and curation}

\paragraph{Matching and curation.} Each structure pair matches an experimental structure of the Crystallography Open Database (COD) with the PBE or PBE+$U$ relaxation of a material of the same reduced formula in MP, the run type being the one MP's mixing rule assigns to the composition. The experimental structures are ordered and inorganic, measured at ambient pressure, and not marked by the COD as erroneous, retracted or duplicate. A COD structure and an MP material match when their primitive cells hold the same number of sites and agree as one structure type up to a uniform scale (pymatgen's StructureMatcher with length, site and angle tolerances of $0.2$, $0.3$ and $5^\circ$); among several matching materials the one of lowest energy is taken. Of the $28{,}798$ matched reports, $25{,}957$ are within the bounds below; of these, $24{,}661$ were measured at no more than $320$~K and $200$~kPa or carry no recorded condition, and $23{,}769$ have the space group of the calculated structure (as reported by the COD or found by spglib at a tolerance of up to $0.1$~\AA). Where several reports match one material, one is kept by criteria that use none of the errors: a temperature between $250$ and $320$~K or none recorded, then the most recent publication, then the lowest $R$ factor, then the lowest COD number. This leaves the $\Nbench$ pairs of the benchmark.

\paragraph{Bounds.} The benchmark contains the $\Nbench$ curated pairs whose DFT volume per atom is within $10\%$ of the
experimental one and whose DFT shape error $s=\|\shp(\bE^*)\|_F$ is at most $10\%$, the same tolerance. A shape error of
$10\%$ is far outside the range of PBE errors of a correctly matched structure, and the bound refers to the lattice
correspondence of the target (Supplementary Section~\ref{si:labels}). Both bounds need the experimental structure: they
define the scope of the reported accuracy and cannot screen new calculations.

\begin{table}[ht]
\centering
\caption{Composition of the benchmark. The $\Nbench$ structure pairs by crystal system and dimensionality of the
bonded network of the DFT structure (0D, molecular; 1D, chains; 2D, layers; 3D, bonded in three dimensions), by
functional, with the number of distinct reduced formulas, the pairs that contain hydrogen, the neutron structures
among them (Supplementary Section~\ref{si:labels}) and the pairs whose atoms correspond.}
\label{tab:si_pop}
\begin{tabular}{ccccccc}
\toprule
triclinic & monoclinic & orthorhombic & tetragonal & trigonal & hexagonal & cubic\\
\midrule
\multicolumn{7}{c}{}\\[-10pt]
\tabcs\\
\midrule
 & 0D & 1D & 2D & 3D & & \\
\midrule
 & $\NzeroD$ & $\NoneD$ & $\NtwoD$ & $\NthreeD$ & & \\
\midrule
PBE & PBE+$U$ & formulas & with H & neutron & atoms correspond & \\
\midrule
$\Npbe$ & $\NpbeU$ & $\Nformula$ & $\NH$ & $\Nneutron$ & $\Nposok$ & \\
\bottomrule
\end{tabular}
\end{table}

\paragraph{Splits.} The ten reported splits ($p=42,\dots,51$) are random partitions with seed $p$ (sizes in Supplementary
Section~\ref{si:results}), with one restriction: $\Nreserved$ pairs were reserved for model selection before the
splits were drawn, so each test set is drawn from the other $\Neligible$. In total, $\Ntested$ distinct pairs are in at
least one test set, and $\Nscored$ of them are evaluated (the others appear in OMat24; Supplementary
Section~\ref{si:overlap}). The $\Ndev$ pairs in no test set, those reserved and those that no split drew, form the
model-selection set (Supplementary Section~\ref{si:dev}).

\section{Input features}
\label{si:inputs}
\subsection*{Definitions}

All inputs are computed from the calculated structure. DFT structures also carry a flag $\chi_U$, $1$ for DFT+$U$
calculations (with the $U$ values of MP~\cite{jain2011_mixing}) and $0$ otherwise, because the descriptors and the
Hessian of MACE-OMAT-0 do not distinguish DFT from DFT+$U$ calculations of the same composition. The flag is $0$ for
all MLIP structures, since an MLIP applies the Hubbard-$U$ convention of its training data by composition.

\emph{Descriptors.} The atom descriptors $\bx_i$ are fixed features of MLIPs pretrained on the Open Materials 2024
(OMat24) data set~\cite{barrosoluque2024_omat24} alone (Supplementary Section~\ref{si:overlap}): the invariant node
features of both layers of MACE-OMAT-0~\cite{batatia2022_mace,batatia2024_macemp0} (medium model), $256$ per atom, and,
for the variant of the cell estimators, the rotation-invariant ($l=0$) channels $\bm\xi_i$ of all ten layers of
eSEN-30M-OMat (equivariant smooth energy network)~\cite{fu2025_esen}, $1{,}280$ per atom, averaged over the structure and
its inversion image (equation~(5) of the main text). Each channel is standardized with the mean and
standard deviation of the training set and, except for the direct strain head, clipped at $\pm8$.

\emph{Hessian.} Let $E_{\rm M}(\bq)$ be the MACE-OMAT-0 energy of the calculated structure deformed by $\bq$
(equation~(1) of the main text). Its Hessian at $\bq=\mathbf 0$ has the blocks
\begin{equation}
\tilde\bH=\begin{pmatrix}V\mathbf C^0&\tilde\bH_{u\gamma}^{\mathsf T}\\ \tilde\bH_{u\gamma}&\bm\Phi\end{pmatrix},
\label{eq:H}
\end{equation}
with $V\mathbf C^0=\partial^2E_{\rm M}/\partial\gam^2$, $\mathbf C^0$ the clamped-ion elastic tensor (Voigt notation,
engineering shear), the internal-strain coupling~\cite{wu2005_relaxedion}
$\tilde\bH_{u\gamma}=\partial^2E_{\rm M}/\partial\bu\,\partial\gam$ of displacements and strain, and the force constants
$\bm\Phi=\partial^2E_{\rm M}/\partial\bu^2$, in eV, eV\,\AA$^{-1}$ and eV\,\AA$^{-2}$, all by automatic differentiation of
MACE-OMAT-0 in double precision. Unstable and zero modes are removed in the coordinates $\mathbf S^{-1}\bq$, with
$\mathbf S=\mathrm{diag}\big(V^{-1/3}\mathbf G_F^{-1/2},\,\I_{3n}\big)$, whose strain part is $V^{1/3}$ times the Kelvin
vector $(E_{11},E_{22},E_{33},\sqrt2E_{23},\sqrt2E_{13},\sqrt2E_{12})$~\cite{mehrabadi1990_kelvin}, so that all blocks
share the unit eV\,\AA$^{-2}$ and a rotation acts orthogonally; in engineering Voigt coordinates the truncation would
depend on the orientation (relative change of the response up to $0.5$ under a rotation for structures with unstable
modes, against $<8\times10^{-14}$ in Kelvin coordinates). With the eigenpairs $(\lambda_k,\mathbf v_k)$ of
$\mathbf S\tilde\bH\mathbf S$, the Hessian of the fits is that of equation~(6) of the main text, and
$\bH^{+}=\mathbf S\big(\sum_{\lambda_k>\eta}\lambda_k^{-1}\mathbf v_k\mathbf v_k^{\mathsf T}\big)\mathbf S$ inverts it on
the retained modes; unstable modes occur in $\negshare\%$ of the structures. The relaxed-ion elastic
tensors~\cite{wu2005_relaxedion}
$\mathbf C=\mathbf C^0-\tilde\bH_{u\gamma}^{\mathsf T}\bm\Phi^{+}\tilde\bH_{u\gamma}/V$, with $\bm\Phi^{+}$ the inverse of
$\bm\Phi$ on its modes above $\eta$, agree with the DFT tensors of MP~\cite{dejong2015_elastic} for the $\Nelastic$
materials with positive-definite tensors in both: Spearman correlation $\Krho$ and median ratio $\Kratio$ for the
Voigt--Reuss--Hill~\cite{hill1952} bulk modulus, $\Grho$ and $\Gratio$ for the shear modulus. The DFT structures are not
minima of MACE-OMAT-0: at them, the median residual stress is $\sigres$~GPa (Frobenius norm) and the median largest
force component $\fres$~eV\,\AA$^{-1}$.

\emph{Relaxation strains.} The calculated structure is relaxed in its space group with MACE-OMAT-0, once without and
once with the D3 dispersion correction~\cite{grimme2010_d3,grimme2011_bj} (settings below). With $\bG_0$ and $\bG_1$
the metric tensors of the cells relaxed without and with D3,
\begin{equation}
\bF_m=\tfrac12\big(\bL^{-1}\bG_m\bL^{-\mathsf T}-\I\big)\quad(m=0,1),\qquad \bT_1=\bF_1,\qquad \bT_2=\bF_1-\bF_0 ,
\label{eq:relax}
\end{equation}
so that $\bT_1$ is the strain to the minimum with dispersion and $\bT_2$ the change of the relaxation strain when
dispersion is added. A mask $\mu_k$ is $1$ where the relaxations behind $\bT_k$ converged and kept the space group of the calculated
structure, and $0$ with $\bT_k=\mathbf 0$ otherwise.

\subsection*{Relaxation settings}

The two relaxations use MACE-OMAT-0 alone and with the D3 correction (Becke--Johnson damping, PBE parameters, two-body
terms within $40$~bohr). The atomic positions are kept in the space group of the calculated structure (tolerance
$10^{-3}$~\AA) and the cell is relaxed with a filter in logarithmic strain coordinates, by at most $800$ steps of the fast inertial relaxation engine
(FIRE)~\cite{bitzek2006_fire} in
single precision to residual forces of $0.01$~eV\,\AA$^{-1}$ and, where that limit is reached, at most $2{,}000$ further
steps in double precision; a relaxation whose space group changes is masked. $\bT_1$ is defined for $\covFone\%$ and
$\bT_2$ for $\covD\%$ of the pairs; relaxations of molecular crystals without dispersion converge least often, since
their energy surface is flat without it.

Where it is defined, $\bT_2$ mainly changes the volume of 3D-bonded structures (medians of $|\tr\bT_2|$ and $\|\dev\bT_2\|_F$ of
$\dTVthree$ and $\dTSthree\%$) and has a larger shape part in layered structures ($\dTVtwo$ and $\dTStwo\%$), chains
($\dTVone$ and $\dTSone\%$) and molecular crystals ($\dTVzero$ and $\dTSzero\%$). Its shape part points towards that of
the target in each class (on the model-selection set, median cosines between $\dev\bT_2$ and $\dev\bE^*$ of
$\dTcostwo$, $\dTcosthree$, $\dTcosone$ and $\dTcoszero$ for layered, 3D-bonded, chain and molecular structures), but its
size is not that of the error (median ratio of the norms $\dTsizetwo$ for layered and $\dTsizethree$ for 3D-bonded
structures); the relaxation strains therefore enter as inputs with learned weights. D3 enters only through these
relaxations: the Hessian and the descriptors are computed without it, and neither the DFT nor the MLIP structures include
a dispersion correction.

\section{Networks, estimators, training and hyperparameters}
\label{si:train}
\subsection*{Atom and pair networks}

Each estimator has its own multilayer perceptrons (MLPs; Fig.~3a of the main text) on invariant inputs,
with the embeddings $\bh_i$, their mean $\bar\bh$ and the pairs $p$ of the main text. Distances enter through $16$
Gaussians $\varphi_k(d)=\exp[-(d-c_k)^2/(0.5~\text{\AA})^2]$, with centres $c_k$ from $0.5$~\AA\ to $r_c$, and the cutoff
$f_c(d)$. Eight learned traceless tensors ($\nu=1,\dots,8$, a channel index) describe the environment of each atom,
\begin{equation}
\mathbf Q^\nu_i=\sum_{p\in\mathcal P_i}f^\nu_p\,\dev(\hatr_p\otimes\hatr_p),\qquad
f^\nu_p=\mathrm{MLP}_\nu\big(\bh_i+\bh_j,\,\bh_i\odot\bh_j,\,\bm\varphi(d_p)\big)\,f_c(d_p),
\label{eq:Q}
\end{equation}
with $\odot$ the elementwise product. With the projections $\vartheta^\nu_{ip}=\hatr_p\mathbf Q^\nu_i\hatr_p^{\mathsf T}$
and $\vartheta^\nu_{jp}=\hatr_p\mathbf Q^\nu_j\hatr_p^{\mathsf T}$ on the direction of pair $p$, a pair network maps the
invariant features
\begin{equation}
\begin{split}
\mathbf z_p=\Big(&\bh_i+\bh_j,\ \bh_i\odot\bh_j,\ \bm\varphi(d_p),\ \bar\bh,\ \chi_U,\\
&\big\{\vartheta^\nu_{ip}+\vartheta^\nu_{jp},\ \vartheta^\nu_{ip}\vartheta^\nu_{jp},\ \tr(\mathbf Q^\nu_i\mathbf Q^\nu_j),\
\tfrac12\hatr_p(\mathbf Q^\nu_i\mathbf Q^\nu_j+\mathbf Q^\nu_j\mathbf Q^\nu_i)\hatr_p^{\mathsf T}\big\}_{\nu=1}^{8},\\
&\big\{100\,\mu_k\hatr_p\bT_k\hatr_p^{\mathsf T},\ 100\,\mu_k\tr\bT_k,\ \mu_k\big\}_{k=1}^{2}\Big)
\end{split}
\label{eq:zp}
\end{equation}
to two outputs $(o_1,o_2)$, which set the predicted Green strain of $d_p$ and its weight as in equation~(7) of the main text. Both are invariant and symmetric in the two atoms of the pair, $w_p$ is positive and vanishes smoothly at $r_c$, and the factor
$10^{-2}$ maps outputs of order one to changes of about $1\%$. The direct strain head uses bond weights
$w^{\rm b}_p=\mathrm{MLP}(\bh_i+\bh_j,\bh_i\odot\bh_j,\bm\varphi(d_p))\,f_c(d_p)$ instead. Each MLP has one or two
hidden layers, of the widths in Supplementary Tables~\ref{tab:si_hyper} and~\ref{tab:si_hyper_mlip}, with sigmoid linear unit (SiLU) activations; the networks of one
estimator, of the atom fit or of the bond-tension network have $0.05$ to $1.31$ million parameters.

\subsection*{Further details of the strain estimators}
The weights $(\alpha_k,\beta_k)$ of the
relaxation strains in the direct strain head (equation~(8) of the main text) start at zero, so that its
training begins without them.

The missing energy $\Delta E$ of the Hessian-regularized DLS fit is an effective free energy: it contains the error of the
functional or potential and, because the calculation is static and the structures were measured at finite temperature,
zero-point and thermal contributions. For $\kappa_0\to\infty$ the minimum of equation~(12) of the main
text approaches the DLS fit; for small $\kappa_0$ it tends, on the modes that $\bH$ retains, to the linear response
$\hat\bq\approx\kappa_0\bH^{+}\mathbf b$ to the forces of the missing energy, which is largest along the soft modes;
along the modes removed from $\bH$ only the restraints act. Because $s_p$ and $w_p$ are computed once, at the calculated
structure, $\Delta E$ is a quadratic model of the missing energy near that structure, not a transferable energy surface,
and it has no temperature input.

In estimator~5 (equation~(13) of the main text), the weights $\lambda_{V,S}=\kappa\big(\sum_pw_p\big)e^{z_{V,S}}$ and the
strength $\kappa=\kappa_0e^{z_K}$ follow from $(z_V,z_S,z_K)=4\tanh(\mathbf o/4)$, where the output
$\mathbf o=\mathrm{MLP}(\bar\bx,\chi_U)$, from the structure mean $\bar\bx$ of the standardized descriptors,
is zero at initialization. Large $z_{V,S}$ pull the volume or the shape towards the direct strain head, and towards
their bound of $-4$ the pull weakens to $e^{-4}\approx2\%$ of its weight at initialization. Equation~(13) of the main text
contains the other estimators as limits: $\bm\Lambda_{\rm d}=\mathbf 0$ with $\kappa=\kappa_0$ gives estimator~4,
$\kappa\to\infty$ with $\bm\Lambda_{\rm d}/\kappa\to\mathbf 0$ the DLS fit of estimator~3, and with $\bu=\mathbf 0$ that
of estimator~2, and $\bm\Lambda_{\rm d}\to\infty$ the strain of the direct strain head, as in estimator~1. The bounded
$z$ keep estimator~5 between these limits.

\subsection*{Training and model selection}

Every rotation-invariant quadratic form of a traceless symmetric error $\Delta\shp=\shp(\hat\bE)-\shp(\bE^*)$ and a
scalar error $\Delta\ln V=\ln(\hat V/V^*)$ combines $\|\Delta\shp\|_F^2$ and $(\Delta\ln V)^2$ (Schur's lemma), so that the cell loss of a structure has one weight $w_0$
between volume and shape,
\begin{equation}
\mathcal L_{\rm cell}=\rho_\delta\big(\|\Delta\shp\|_F^2+\tfrac{w_0}{3}(\Delta\ln V)^2\big),\qquad
\rho_\delta(x)=\delta^2\big(\sqrt{1+x/\delta^2}-1\big),
\label{eq:loss}
\end{equation}
where $w_0=1$ gives $\|\Delta\shp+\tfrac13\Delta\ln V\,\I\|_F^2$. An output whose $\I+2\hat\bE$ has an eigenvalue at or
below $10^{-6}$, for which the error measures are undefined, takes the quadratic form of the same loss at its target
instead (the first-order expansion of the error measures at $\bE^*$).
The pseudo-Huber function $\rho_\delta$ is quadratic below a residual $\delta$, a
strain of $0.24\%$, and linear above it, so that the loss is close to the mean absolute strain error; $w_0$ and
$\delta$ come from a search ($3.35$ and $2.36\times10^{-3}$ for DFT; see \emph{Loss settings and restraint strength} below). The estimators with displacements add an atom
loss,
\begin{equation}
\mathcal L_{\rm at}=\rho_{\delta_{\rm at}}\Big(\frac{\sum_b\omega_be_b^2}{\sum_b\omega_b}
+w_u\frac{\sum_{k\notin\mathrm H}\|\hat\bu_k-\bu^*_k\|^2}{n'\bar d^{\,2}}\Big),
\label{eq:lossat}
\end{equation}
with the sums over the first-shell bonds $b$ of the experimental structure, selected as for $\mathcal B$ but
including hydrogen, $e_b=\mathbf a_b\cdot(\hat\bq-\bq^*)$ the linearized relative error of the length of bond $b$
(equation~(9) of the main text, with the length and direction of the bond in the calculated structure), $\bq^*=(\gam(\bE^*),\bu^*)$,
$\bar d$ the mean length of the first-shell bonds of the calculated structure, $\omega_b=1$ for bonds between atoms
other than hydrogen and $0$ for bonds to hydrogen (both terms vanish for pairs whose atoms do not correspond), and $w_u$ and
$\delta_{\rm at}$ set for each estimator. Bonds to hydrogen and the
displacements of hydrogen atoms carry no weight, so the hydrogen positions enter the training of the estimators and the atom fit only through the mean removed in
equation~(3) of the main text and the check that the atoms correspond; the cells of all pairs and the other atoms of the
$\NH$ hydrogen-containing pairs enter in full. The atom loss uses the
displacements at the experimental cell (equation~(15) of the main text with $\gam=\gam(\bE^*)$). Estimator~5 differs: it
applies $\rho_\delta$ separately to the shape and volume terms of equation~\eqref{eq:loss}, and $\rho_{\delta_{\rm at}}$
to the bond and the translation-free displacement terms of equation~\eqref{eq:lossat}, evaluates the atom loss at
the structure it predicts and adds the cell losses of its head and of its Hessian-regularized DLS fit without the
prior.
Gradients of the losses pass through the linear solves of the fits. Each estimator is trained separately for each split, from three
random initializations, with AdamW~\cite{loshchilov2019} and early stopping on the validation loss (for estimator~5 on its
cell part; settings in Supplementary Table~\ref{tab:si_hyper}).

\emph{Model selection.} Design choices were compared on ten folds of the model-selection set (Supplementary Section~\ref{si:dev}), and
hyperparameters and $\kappa_0$ were set by searches on further partitions of it into five folds; each MLIP correction keeps the
design chosen for DFT and repeats the searches on its own structures. Paired fold differences $y_j$ are compared with the
corrected resampled $t$-test~\cite{nadeau2003},
\begin{equation}
t=\frac{\bar y}{\sqrt{(1/J+n_2/n_1)\,s_y^2}},
\label{eq:ttest}
\end{equation}
with $\bar y$ and $s_y^2$ the mean and variance of the $y_j$, $n_2$ and $n_1$ the numbers of test and training pairs per
fold and $J-1$ degrees of freedom; its second term accounts for the overlap of the training sets between folds. Holm's
correction~\cite{holm1979} accounts for the number of options in a comparison. The tests were used only to choose
between designs; the accuracy of the model is estimated on the test sets and the grouped folds.

\subsection*{Bond-tension network}

The ablation of the Hessian (Supplementary Section~\ref{si:results}) uses a bond-tension network on the MACE-OMAT-0
descriptors, without the relaxation strains. Its
pair network gives each pair within $r_c$ a tension
$t_p=10^{-2}\,\mathrm{MLP}\big(\bh_i+\bh_j,\,\bh_i\odot\bh_j,\,\bm\varphi(d_p)\big)\,f_c(d_p)$ (eV\,\AA$^{-1}$; $t_p>0$
pushes the two atoms apart), whose forces on the atoms are $\bg=\sum_pt_p\,\partial d_p/\partial\bu$. At a cell
$\gam$ they move the atoms through the Hessian,
\begin{equation}
\bu^{\rm B}(\gam)=\mathbf K_\zeta\bg+c\,\bW_3\bm\Gamma_\zeta\gam,\qquad
\mathbf K_\zeta=(\I+\zeta\mathbf K)^{-1}\mathbf K,\qquad \bm\Gamma_\zeta=(\I+\zeta\mathbf K)^{-1}\bm\Gamma,
\label{eq:hB}
\end{equation}
with $\mathbf K$ and $\bm\Gamma$ the clamped-cell compliance and the relaxed-ion internal strain of the Hessian of
equation~(6) of the main text ($\bm\Phi^{+}$ and $-\bm\Phi^{+}\tilde\bH_{u\gamma}$ for a structure without
unstable modes), $\zeta$ (eV\,\AA$^{-2}$) a damping of the soft internal modes,
$c=\min\big(1,\,0.5~\text{\AA}/\max_k\|(\bW_3\bm\Gamma_\zeta\gam)_k\|\big)$ a cap on the internal strain, and
$\bW_3=\bW\otimes\I_3$ the riding of the hydrogen atoms: $W_{kj}=1/|\mathcal N_k|$ for $j\in\mathcal N_k$, the
first-shell neighbours other than hydrogen of a hydrogen atom $k$ in the calculated structure that lie within $10^{-3}$~\AA\ of the
nearest one, and $W_{kj}=\delta_{kj}$ for the other atoms and for hydrogen atoms without such neighbours. The network is trained
at the experimental cell with the loss of equation~\eqref{eq:lossat}, with its own $w_u$ and $\delta_{\rm at}$, in which
the bonds to hydrogen of the neutron structures also carry weight ($\omega_b=1$).

\subsection*{Software and versions}

The networks, fits and losses are implemented in PyTorch~\cite{paszke2019_pytorch} (version 2.11.0 on GH200
GPUs, 2.8.0 on CPUs); MACE-OMAT-0 runs in mace-torch 0.3.16 and the D3 correction in
torch-dftd 0.5.3, and these computations handle structures with ASE 3.26.0~\cite{larsen2017_ase}, pymatgen
2024.8.9~\cite{ong2013_pymatgen} and spglib 2.7.0~\cite{togo2018_spglib}. eSEN-30M-OMat runs in fairchem-core 1.10.0
with PyTorch 2.9.1, ASE 3.29.0 and spglib 2.6.0. The tight relaxations of Supplementary Section~\ref{si:overlap} were made with VASP 5.4.4, and the MLIP structures
come from the released TECE-OAM-RRA-1.0 (version 2026.07.05) and
EquFlashV2-45M-OAM (version 2026.06.11) models, each run in its own environment (tace~$\geq$~0.2.0,
fairchem-core~$\geq$~1.10.0, PyTorch~$\geq$~2.9, ASE~$\geq$~3.29).

\subsection*{Settings and hyperparameters}

The three initializations use PyTorch seeds $0$, $1$ and $2$; Supplementary Table~\ref{tab:si_hyper} lists the settings.
The learning rate and hidden width of each estimator, and the weight $w_u$ and scale $\delta_{\rm at}$ of its atom loss,
were chosen on the model-selection set from $16$ or $32$ Latin-hypercube configurations~\cite{mckay1979} (learning rate
log-uniform in $[10^{-4},3\times10^{-3}]$, width in $\{64,128,256,384\}$, $w_u$ log-uniform in $[0.1,3]$,
$\delta_{\rm at}$ log-uniform in $[0.003,0.05]$), each evaluated out of fold over five folds by $\bar e_S+\tfrac12\bar e_V$
(cell estimators) or by $\bar e_B$ at the experimental cell (atom fit); for DFT, estimators~4 and 5 and the atom fit
were searched at $\kappa_0=30$~eV. The weight of one half favours the shape error, the target of the design choices,
while still penalizing a larger volume error. The bond-tension network was searched in the same
way, with its batch size in $\{32,64,128\}$ and its damping $\zeta$ log-uniform in $[0.1,2]$~eV\,\AA$^{-2}$ added, and
takes, of its eight configurations with the lowest $\bar e_B$, the one with the lowest error of the bonds to hydrogen. The
variant with eSEN-30M-OMat descriptors takes the settings selected for those descriptors (Supplementary
Table~\ref{tab:si_hyper}). The networks of the model and of its variants were trained, and the model-selection runs made, on NVIDIA GH200 GPUs with the same code.

\paragraph{Loss settings and restraint strength.} The weight $w_0$ and the scale $\delta$ of the cell loss
(equation~\eqref{eq:loss}), the weight decay and the batch size come from a search of $\setN$ Latin-hypercube
configurations of a direct strain head on the MACE-OMAT-0 descriptors with the relaxation strains ($w_0$
log-uniform in $[1,5]$, $\delta$ in $[10^{-3},3\times10^{-2}]$, learning rate in $[10^{-4},3\times10^{-3}]$, weight decay
in $[10^{-6},10^{-2}]$, width in $\{64,128,256\}$, batch size in $\{32,64,128\}$), evaluated out of fold over five
folds of the model-selection set by $J=\bar e_S+\tfrac12\bar e_V$. The optimum is flat: $J$ ranges from $\setJlo$ to $\setJhi$, and the
five best configurations, all with $\delta\le\setTopDhi$ and $w_0$ from $\setTopWlo$ to $\setTopWhi$, lie within
$\setJtop$ percentage points of each other. A small $\delta$ makes the loss close to the mean absolute error that $e_V$
and $e_S$ measure, and $w_0>1$ weights the volume more than the Frobenius norm of $\Delta\shp+\tfrac13\Delta\ln V\,\I$
would ($w_0=1$); the same $w_0$ and $\delta$ define the cell loss of each estimator. The restraint strength $\kappa_0$ is
set the same way for each source of structures: the cell takes the value of lowest mean shape error of the
Hessian-regularized DLS fit alone, the atoms the value of lowest mean bond-length error, at the experimental cell, of the
atom fit (five folds of the model-selection set; values within $10^{-4}$ points count as equal, and the smaller $\kappa_0$ is taken; the
DLS limit $\kappa_0\to\infty$, shown in the tables, is not among the values considered). For
DFT the cell sweep used a reduced estimator (MACE-OMAT-0 descriptors, no relaxation strains, width $256$) and the atom
sweep the settings of Supplementary Table~\ref{tab:si_hyper} (Supplementary Table~\ref{tab:si_kappa}).

\begin{table}[ht]
\centering
\small
\caption{Restraint strength $\kappa_0$ of the DFT correction. Five folds of the model-selection set, one initialization; means over
the folds of the fold means (\%). Cell: the Hessian-regularized DLS fit alone, reduced (MACE-OMAT-0 descriptors,
no relaxation strains, width $256$); atoms: the Hessian-regularized DLS fit of the atoms with the settings of
Supplementary Table~\ref{tab:si_hyper}, $e_B$ at the experimental cell. $\Delta$, mean paired difference to the value
used ($^*$), computed before rounding.}
\label{tab:si_kappa}
\begin{tabular}{@{}lccccc@{}}
\toprule
 & \multicolumn{3}{c}{cell} & \multicolumn{2}{c}{atoms}\\
\cmidrule(lr){2-4}\cmidrule(l){5-6}
$\kappa_0$ (eV) & $e_V$ (\%) & $e_S$ (\%) & $\Delta e_S$ & $e_B$ (\%) & $\Delta e_B$\\
\midrule
\tabsikappa
\bottomrule
\end{tabular}
\end{table}

\begin{table}[ht]
\centering
\small
\caption{Restraint strength $\kappa_0$ of the MLIP corrections. As Supplementary Table~\ref{tab:si_kappa}, on five
folds of the model-selection set of each MLIP's structures; cell: the Hessian-regularized DLS fit alone with the MACE-OMAT-0
descriptors and the relaxation strains, with the learning rate and width of the configuration of each MLIP's own loss-settings search;
atoms: the Hessian-regularized DLS fit of the atoms with the settings of Supplementary
Table~\ref{tab:si_hyper_mlip}.}
\label{tab:si_kappa_mlip}
\begin{tabular}{@{}lccccc@{}}
\toprule
 & \multicolumn{3}{c}{cell} & \multicolumn{2}{c}{atoms}\\
\cmidrule(lr){2-4}\cmidrule(l){5-6}
$\kappa_0$ (eV) & $e_V$ (\%) & $e_S$ (\%) & $\Delta e_S$ & $e_B$ (\%) & $\Delta e_B$\\
\midrule
\tabsikappamlip
\bottomrule
\end{tabular}
\end{table}

\begin{table}[ht]
\centering
\small
\caption{Training settings of the DFT correction. Learning rate, hidden width $H$, atom-loss parameters and
wall-clock limit per split (Supplementary Table~\ref{tab:si_hyper_mlip} for the MLIP corrections). Each network except the bond-tension network$^b$ uses AdamW~\cite{loshchilov2019} with weight decay $1.68\times10^{-6}$, batches of $128$ structures, at most
$300$ epochs, the learning rate halved after $10$ and training stopped after $30$ epochs without improvement of the
validation loss (of its cell part for the Hessian-regularized DLS fit with a strain prior), gradients clipped at norm
$10$, and the cell loss of equation~\eqref{eq:loss} with $\delta=2.36\times10^{-3}$ and
$w_0=3.35$ (separate volume and shape terms for the Hessian-regularized DLS fit with a strain prior). $w_u$ and $\delta_{\rm at}$: weight of the displacement term and scale of the atom loss
(equation~\eqref{eq:lossat}); $\kappa_0$: restraint strength. Each network except those of the
direct strain head and the bond-tension network uses the eight tensors $\mathbf Q^\nu_i$ (equation~\eqref{eq:Q}). Each network uses the
MACE-OMAT-0 descriptors and, except the bond-tension network, the relaxation strains. $^a$Restraint strength set per structure; volume and shape
losses separate. $^b$Batches of $128$ structures, weight decay $10^{-4}$, at most $200$ epochs, the learning rate halved
after $8$ and training stopped after $25$ epochs without improvement, gradients clipped at norm $10$; $\zeta$: damping
(equation~\eqref{eq:hB}). Bottom: the settings selected for the variant with eSEN-30M-OMat descriptors, for the four of its five cell estimators whose settings differ from those above.}
\label{tab:si_hyper}
\begin{tabular}{@{}llllll@{}}
\toprule
Estimator & rate & $H$ & $w_u$ & $\delta_{\rm at}$ & other\\
\midrule
direct strain head & $2.55\times10^{-3}$ & 384 & -- & -- & 1.5~h\\
DLS fit of the cell & $1.01\times10^{-3}$ & 384 & -- & -- & 2~h\\
DLS fit of the cell and atoms & $5.72\times10^{-4}$ & 64 & $0.465$ & $3.70\times10^{-3}$ & 2~h\\
Hessian-regularized DLS fit & $4.81\times10^{-4}$ & 128 & $0.409$ & $6.31\times10^{-3}$ & $\kappa_0=\kapK$~eV, 2~h\\
with a strain prior & $5.26\times10^{-4}$ & 128 & $2.36$ & $4.58\times10^{-3}$ & $\kappa_0=\kapK$~eV, adaptive$^a$, 2~h\\
Hessian-regularized DLS fit, atoms & $1.13\times10^{-3}$ & 384 & $1.69$ & $4.03\times10^{-3}$ & $\kappa_0=\kapBK$~eV, 2~h\\
bond-tension network$^b$ & $9.22\times10^{-4}$ & 256 & $0.618$ & $3.95\times10^{-3}$ & $\zeta=1.29$~eV\,\AA$^{-2}$\\
\midrule
direct strain head & $1.16\times10^{-4}$ & 384 & -- & -- & \\
DLS fit of the cell & $1.74\times10^{-4}$ & 384 & -- & -- & \\
DLS fit of the cell and atoms & $4.92\times10^{-4}$ & 128 & $2.15$ & $3.49\times10^{-3}$ & \\
Hessian-regularized DLS fit & $1.73\times10^{-4}$ & 384 & $0.485$ & $4.22\times10^{-3}$ & \\
\bottomrule
\end{tabular}
\end{table}

The auxiliary cell losses of the Hessian-regularized DLS fit with a strain prior, those of its direct head alone and
of its fit without the prior, keep both parts accurate when the prior weights favour the other.

The regularization of equation~(10) of the main text uses $\lambda_g=10^{-4}\sum_pw_p$ and
$\lambda_u=10^{-3}\cdot2\sum_pw_p/d_p^2/(3n)$. Each solve is checked by its
normwise backward error ($\leq10^{-10}\sqrt{N}$ for $N$ unknowns) and repeated on the CPU if the check
fails.

\paragraph{MLIP corrections.} Each MLIP correction keeps the folds, inputs and estimators of the DFT correction and is
tuned on the model-selection set of its own structures. The same $\setN$ loss-settings configurations, run with the final
inputs, give $w_0=\mlipWtece$ and $\mlipWeqf$, $\delta=\mlipDtece$ and $\mlipDeqf$, weight decays of $\mlipWDtece$ and $\mlipWDeqf$ and batches of $\mlipBStece$ and $\mlipBSeqf$
structures for TECE-OAM-RRA-1.0 and EquFlashV2-45M-OAM. The learning rate, width and atom-loss parameters of each
estimator were searched with the configurations of the DFT search (Supplementary Table~\ref{tab:si_hyper_mlip}), at
restraint strengths chosen by a criterion that also weighted the volume error ($100$~eV for TECE-OAM-RRA-1.0 and $30$~eV for
EquFlashV2-45M-OAM); with these settings, the choice described above gives $\kappa_0=\mlipKtece$~eV for the cells of both corrections
(sweeps with the relaxation strains) and $\mlipKBtece$~eV for their atoms (Supplementary
Table~\ref{tab:si_kappa_mlip}).

\begin{table}[ht]
\centering
\footnotesize
\setlength{\tabcolsep}{3pt}
\caption{Training settings of the MLIP corrections. Chosen on the model-selection set of each MLIP's structures:
learning rate, hidden width $H$, atom-loss weight $w_u$ and scale $\delta_{\rm at}$; the other settings as in
Supplementary Table~\ref{tab:si_hyper}, with the loss settings, weight decay, batch size and $\kappa_0$ of the text.}
\label{tab:si_hyper_mlip}
\begin{tabular}{@{}lcccccccc@{}}
\toprule
 & \multicolumn{4}{c}{TECE-OAM-RRA-1.0} & \multicolumn{4}{c}{EquFlashV2-45M-OAM}\\
Estimator & rate & $H$ & $w_u$ & $\delta_{\rm at}$ & rate & $H$ & $w_u$ & $\delta_{\rm at}$\\
\midrule
\tabsihypermlip
\bottomrule
\end{tabular}
\end{table}

\section{Prediction for new structures}
\label{si:new}

A new structure needs only its calculated geometry and, for DFT, whether DFT+$U$ was used. Its descriptors, exact
Hessian, pair lists and two relaxations are computed, and the networks of all ten splits, $180$ in all, are applied
with the standardization of their training data; their strains and displacements are averaged.

\paragraph{Outputs.} Besides the corrected cell $\hat{\bL}=\bL(\I+2\hat\bE)^{1/2}$ and fractional coordinates
$\hat{\mathbf f}_k=\mathbf f_k+\hat\bu_k\bL^{-1}$, a prediction reports two spreads of the five cell estimates:
$s_S=\big(\tfrac15\sum_r\|\mathrm{dev}(\hat\bE_r-\hat\bE)\|_F^2\big)^{1/2}$, the root-mean-square deviation of their
traceless parts from the mean, and $s_V=\mathrm{std}_r\,\tr\hat\bE_r$, the standard deviation of their volume strains. A mean strain without a valid cell (an eigenvalue of $\I+2\hat\bE$ at or
below $10^{-6}$) returns the uncorrected structure. The volume spread $s_V$ is the weaker indicator: its Spearman correlation with the
corrected volume error is $\Tsprrhovol$, against $\Tsprrho$ between $s_S$ and the corrected shape error (main text).

\section{Pretraining data, overlap with the targets and relaxation history}
\label{si:overlap}
\subsection*{Overlap with pretraining data}

\paragraph{MPtrj.} Potentials trained on MPtrj~\cite{deng2023_chgnet} have seen about half of the experimental structures: a
structure match of each MPtrj frame of the benchmark materials against them (tolerances $0.2\%$ in the lattice lengths,
$0.1^\circ$ in the angles and $0.01\,(V/n)^{1/3}$ in the positions, supercells allowed) finds a frame that is the
experimental structure for $52\%$ of the pairs. Such potentials, among them MACE-MP, MACE-MPA-0,
TECE-OAM-RRA-1.0~\cite{tece2026}, EquFlashV2-45M-OAM and the models trained jointly on OMat24, its subsample of the
Alexandria structures~\cite{schmidt2023_alexandria} (sAlex) and MPtrj, therefore supply no input; TECE-OAM-RRA-1.0 and
EquFlashV2-45M-OAM appear only as methods to be corrected (Supplementary Table~\ref{tab:si_exposure}).

\paragraph{OMat24.} The input potentials are trained on OMat24 alone~\cite{barrosoluque2024_omat24}, whose
configurations are perturbed from DFT-relaxed Alexandria structures and include no relaxation started from an
experimental structure. Matching all $101.8$ million frames of its $11$ subsets against the experimental structures with
the same tolerances finds $44$ pairs, most of them in molecular-dynamics runs started from DFT minima; for $30$ of them the
DFT structure itself lies within the tolerances of the experimental one. Looser tolerances ($0.5\%$, $0.5^\circ$,
$0.05\,(V/n)^{1/3}$), with the frame nearer to the experimental than to the DFT structure, add $80$ more. These $\Nomat$ pairs, all in the benchmark, stay in training but are left out of all reported errors. Two indirect links remain:
the Boltzmann-sampled subsets of OMat24 were selected with the energies of a model trained on MPtrj, and the Alexandria
minima were relaxed from structures whose sources include experimental databases.

\subsection*{Relaxation history of the inputs}

\paragraph{Tight relaxations.} Most MP relaxations stop at an energy criterion, so a DFT input may retain part of its
starting geometry, which the correction could exploit. To test this, we took the DFT inputs of a random sample of $\memNR$ pairs of the model-selection set with at most
$16$ atoms and relaxed them again with VASP~\cite{kresse1996_prb,kresse1996_cms}. The relaxations used the settings of MP (PBE or PBE+$U$ as assigned by
the MPRelaxSet of pymatgen~\cite{ong2013_pymatgen}, a plane-wave cutoff of $520$~eV and projector augmented-wave
potentials~\cite{blochl1994_paw,kresse1999_paw} of the PBE.52 release) and ran to forces of $0.005$~eV\,\AA$^{-1}$
in at least two consecutive relaxations. Of the $\memNR$, $\memConvR$ converged. Their tight minima
differ from the inputs by a median of $\memRdevmed\%$ in shape, $\memRtrmed\%$ in volume and less than $0.001$~\AA\
in the atomic positions, and for the $\memIn$ of them whose experimental structure is not in OMat24 their DFT shape
error is larger ($\memRSdfttight$ against $\memRSdftinp\%$) at a similar volume error. The correction of the grouped folds, whose models
did not train on these pairs, removes a similar share of the shape error from the tight minima as from the inputs
($\memRSredtight$ against $\memRSredinp\%$), although its errors are higher there, by $\memIVrel$, $\memISrel$ and
$\memIBrel\%$ in volume, shape and bond length. The correction is thus calibrated to relaxations made with the MP
protocol, as in the databases it is meant for.

\providecommand{\bibcommenthead}{}
\bibliographystyle{sn-nature}
\bibliography{references}